\documentclass[a4paper,fleqn,usenatbib]{mnras}

\usepackage{newtxtext,newtxmath}

\usepackage[T1]{fontenc}
\usepackage{ae,aecompl}

\usepackage{graphicx}	
\usepackage{amsmath}	
\usepackage{amssymb}	

\newcommand\resp[1]{{\textcolor{black}{#1}}} 

\usepackage{bm}
\newcommand\vct[1]{\bm{#1}}

\title[Photo-$z$ of blended sources]{Bayesian photometric redshifts of blended sources}

\author[D. M. Jones \& A. F. Heavens]{
	Daniel M. Jones,$^{1}$\thanks{E-mail: d.jones15@imperial.ac.uk}
	Alan F. Heavens,$^{1}$
	\\
	$^{1}$Astrophysics Group \& Imperial Centre for Inference and Cosmology, Imperial College London, London SW7 2A\\
}

\date{Accepted XXX. Received YYY; in original form ZZZ}

\pubyear{2018}

\begin{document}
\label{firstpage}
\pagerange{\pageref{firstpage}--\pageref{lastpage}}
\maketitle

\begin{abstract}
Photometric redshifts are necessary for enabling large-scale multicolour galaxy surveys to interpret their data and constrain cosmological parameters.  
While the increased depth of future surveys such as the Large Synoptic Survey Telescope (LSST) will produce higher precision constraints, it will also increase the fraction of sources that are blended.
In this paper, we present a Bayesian photometric redshift method for blended sources with an arbitrary number of intrinsic components.
This method generalises \resp{existing template-based Bayesian photometric redshift (BPZ) methods}, and produces joint posterior distributions for the component redshifts that allow uncertainties to be propagated in a principled way. 
Using Bayesian model comparison, we infer the probability that a source is blended and the number of components that it contains. 
\resp{We extend our formalism to the case where sources are blended in some bands and resolved in others. Applying this to the combination of LSST- and Euclid-like surveys, we find that the addition of resolved photometry results in a significant improvement in the reduction of outliers over the fully-blended case.}
We make available \texttt{blendz}, a Python implementation of our method.
\end{abstract}




\section{Introduction}

Current photometric surveys such as CFHTLens~\citep{cfhtData}, KiDS~\citep{kidsData} and DES~\citep{desData} image galaxies over large volumes of the Universe to probe the growth of structure and the distribution of matter on large scales. Through techniques such as galaxy clustering and cosmic shear, these surveys are able to constrain cosmological parameters and conduct tests of the standard $\Lambda$CDM cosmological model~\citep[e.g.,][]{cfhtConstraints, desConstraints}. 

These observational tests require the redshift distribution of the sample to make model predictions for comparison. Additional information can also be obtained by considering the redshift dependence using tomography, where galaxies are placed into one of several redshift bins~\citep[e.g.][]{tomographyHu, tomographyConstraints, tomographyKids}. However, the large number of sources required to constrain cosmological parameters to high precision makes obtaining spectroscopic redshifts for the entire sample unfeasible. As a result, photometric redshifts are a vital part of the cosmological analysis pipeline of galaxy surveys. 

Photometric redshift methods seek to infer the redshift of galaxies from noisy observations of their flux in several broadband filters. They provide an alternative to spectroscopic redshifts that requires less telescope time, at the expense of a reduction in precision. As a result, photometric redshifts can be applied to galaxies too faint and samples too large for spectroscopic observations. There are two general classifications for photometric redshift methods that utilise flux information; template-based and empirical methods. 

Template-based methods use a set of galaxy spectra that are assumed to be representative of every galaxy they are applied to. These templates are redshifted and integrated over the response function of each filter to produce predictions of observed fluxes. These predictions are then used to infer the redshift from the observed fluxes. Maximum likelihood methods~\citep[e.g.,][]{hyperz, lePhare} find the best fitting template by minimising $\chi^2$ to estimate the redshift. Bayesian methods, introduced by \cite{bpz}, marginalise over all templates to produce a posterior redshift distribution. This correctly accounts for the uncertainty in the galaxy template that is ignored by maximum likelihood methods. Bayesian methods also include prior distributions that can reduce catastrophic outliers.

Empirical methods estimate redshifts by fitting for the mapping between flux and redshift from a set of training data, rather than specifying it \textit{a priori} through a template set. This mapping is typically found using machine learning methods such as neural networks~\citep[e.g., ][]{annz, annz2}, Gaussian processes~\citep[e.g.,][]{firstGPPhotoz, newGPz}, and random forests~\citep[e.g., ][]{randomForest, tpz}. These methods are examples of supervised learning; they require large datasets of fluxes and associated spectroscopic redshifts that are representative of the sample they are applied to. 

If representative data are available, empirical methods are typically more accurate than template-based methods~\citep{photozAccuracy}. However, redshift estimates of galaxies not represented by the training data are much less reliable~\citep{mlRedshiftBias}. The common case where spectroscopic training data is shallower than the photometry can lead to biases where the redshifts of high redshift galaxies are underestimated~\citep{mlRedshiftShallow}.

\resp{
In practice, template-based and empirical methods are not so distinct. The priors of Bayesian methods typically include a set of parameters that are fitted using a set of training data~\citep[e.g.,][also see section~\ref{sec:calibrate-priors}]{bpz, schmidtPriors}. In addition, recent applications of photometric redshifts have used hybrid methods that combine a template-based approach with machine learning methods~\citep[e.g.,][]{somsHybrid, hierarchyHybrid, gpAndHybrid}.
}

In addition to these methods, clustering redshift methods~\citep[e.g.,][]{clusterFirst, clusterSchmidt, clusterMenard} cross correlate the angular positions of a photometric sample with a spectroscopic sample to estimate the redshift distribution. Clustering redshift methods do not model fluxes as a function of redshift, instead only using the spatial information of photometric data. As such, clustering redshifts are complementary to other photometric redshift methods; \cite{clusterDES} uses clustering redshifts to calibrate biases from other photometric redshift methods, for example.

\resp{
Ensuring that photometric redshifts are accurate and precise is necessary for obtaining unbiased constraints on cosmological parameters. \cite{hutererSystematics} found that future tomographic surveys would require the mean of each redshift bin to be known to a precision of $0.003$, though this requirement can be reduced by self-calibration~\citep[e.g., ][]{hutererSystematics, calibrationSun, calibrationSamuroff} and combining weak lensing data with other cosmological probes such as baryonic acoustic oscillations~\citep{lsstsciencebook}. Photometric redshifts are also important in the calibration of other systematics. Multiplicative biases in the measurement of shear can be detected and corrected for, provided that photometric redshifts of galaxies in the sample are unbiased~\citep{shapeBiasStudy}. Weak lensing shape measurement biases can themselves also be redshift dependent; without unbiased redshift estimates to make corrections, these can lead to biases of a few percent in the cosmological parameters $\sigma_8$ and $w_0$~\citep{zDependentShapeBias}.
}

{Another} key part of precision cosmology is an accurate understanding of uncertainties in parameter constraints. To enable this, uncertainties arising from each step of the analysis should be accounted for and propagated onwards. In cosmological analyses, this is typically accomplished using a Bayesian framework~\citep[e.g.,][]{kidsResults, desYr1CosmicShear}, allowing these uncertainties to be combined and marginalised over for the final constraints. It is therefore essential that photometric redshift methods provide not only point estimates of redshifts, but also a measure of their uncertainties.

The uncertainty associated with a redshift estimate can be represented by a single number, i.e., a point estimate with an error bar. However, doing this necessitates making an assumption about how the error is distributed.  Uncertainties in photometric redshifts can be highly non-Gaussian, and so are poorly described by a single number such as the variance. Photometric redshift methods that instead characterise their results using a probability distribution function (PDF) can capture all of this information. 

Photometric redshifts can also suffer from degeneracies that result in high-redshift galaxies having similar colours to those at low redshifts~\citep[e.g.,][]{lymanBalmer}. As a result, several well-separated redshifts are plausible, and an accurate representation of the uncertainty should reflect this. While this can be easily described with a multimodal PDF, a single number can be misleading. Error bars that cover the full range of parameter space between the low- and high-redshift estimates do not show that redshifts between these are disfavoured, inflating uncertainties. 

Several photometric redshift methods are able to produce PDFs as their result. Bayesian template-based methods \citep[e.g.,][]{bpz} produce a posterior distribution, a PDF of the model parameters conditioned on the data and any model assumptions. In addition to the galaxy redshift, these model parameters can include other quantities of interest such the galaxy template \citep{zebra}. A joint posterior over all of these parameters contains information about the uncertainty of each, including any correlation between them. Machine learning methods can also produce PDFs by utilising ensemble techniques, where the predictions of several models are combined to produce a distribution. Examples of this technique include the combination of decision trees in a random forest \citep{tpz} and committees of neural networks constructed with different network architectures and initialised randomly \citep{neuralNetPDF}.

Future galaxy surveys such as the Large Synoptic Survey Telescope~\citep[LSST, ][]{lsstSummary} will obtain extremely high precision constraints on cosmological parameters. By utilising deeper photometry, these surveys will probe greater volumes than previously, resulting in an increased number density of galaxies imaged. While this increased depth drives the high precision these surveys will achieve, it also increases the fraction of objects that overlap with others along the line of sight, known as blending~\citep{lsstDensity}. 

Most existing deblending methods do not utilise the colour information from photometry, instead using the spatial information contained in an image from a single band. The commonly used SExctrator~\citep{sextractor} searches for adjacent pixels on a flux-thresholded map that separate into disjoint regions as the threshold is increased. Doing this for many thresholds allows each pixel to be assigned to a single object, contributing the entirety of its flux to that object. The SDSS deblender~\citep{sdssDeblend} lifts this restriction, splitting the flux proportionally between objects based on object templates. These templates are constructed by finding peaks in the image and assuming symmetry around them, comparing pairs of pixels and setting them to be equal. Profile fitting methods~\citep[e.g.,][]{galfit, profit} forward model the image using physical profile models, deblending by directly fitting for the galaxy properties. In far-infrared astronomy, blending is common due to the reduced angular resolution of these instruments compared to optical telescopes. As a result, galaxies that are well resolved in optical observations may become blended in the far-infrared.  Deblending methods designed for this case such as \cite{xid} can use the unblended observations to place strong priors on the number and position of sources. We refer to this mix of blended and unblended observations as \textit{partial blending} throughout this paper.

The ability for most deblending methods to successfully identify blended galaxies depends on their angular separation. Galaxies with too small an angular separation are instead identified as single sources. \cite{lsstBlendReport} estimate that $45-55\%$ of sources in LSST will be blended, with $15-20\%$ of all sources being misidentified as a single source, referred to as ambiguously blended objects.

Blending of sources can have an impact that is significant for constraining cosmological parameters. \cite{blendEllipticities} estimate that ambiguously blended objects in LSST will result in an increase in shear noise of $14\%$ for the deepest photometry ($i < 27$) and $7\%$ for the gold standard sample ($i < 25.3$). Since these ambiguous blends are difficult to separate due to their small angular separation, deblending methods that incorporate colour information could be beneficial. Recent deblending methods such as MuSCADeT~\citep{muscadet} and \textsc{scarlet}~\citep{scarlet} incorporate this colour information by using wavelet transforms, enforcing that the representation of components is sparse in this space.

Deblending methods that produce a set of component-separated maps are useful for later applying existing analysis techniques designed for individual components to. However, splitting the analysis in this way can lose information, such as the correlation between deblending parameters and the parameters in a subsequent analysis. 
An analysis method that jointly constrains parameters directly from blended data provides a self-consistent, principled way to characterise and propagate this information.

In this paper, we present a method that generalises the \cite{bpz} Bayesian photometric redshift (BPZ) method to the case of blended observations. This is a template-based method where the task of determining the component redshifts is cast as a Bayesian parameter inference problem. The product of such an inference is a joint posterior distribution of the redshift and magnitude of each component in the blended source. 
This distribution characterises the complete statistical uncertainty in the result in a way that can be propagated through the rest of the cosmological analysis. 
Determining the number of components in an observed source, i.e., whether or not it is blended, is treated as a model comparison problem. In this way, our method allows the identification of blended sources from aperture photometry alone.

Throughout, we use \textit{source} to refer to the (possibly) blended system that is observed, and \textit{component} to refer to the underlying physical objects that make up this source. For parameters defined for each component in a source, we index over component using greek letters and indicate the collection of these using sets, i.e., $\{\theta\} \equiv \{\theta_\alpha, \theta_\beta, \dots \theta_{N}\}$. Vector quantities defined for each filter band are in bold $\vct{q}$, and observed quantities are denoted with a hat $\hat q$. Where necessary, quantities defined for a specific number of components are distinguished by a subscript number in brackets, i.e., $q^{(1)}$ is the definition of $q$ for a single component. A summary of our notation is provided in Table~\ref{tab:notation}.

\begin{table*}
	\caption{A summary of the notation used throughout this paper.}
	\centering
	\label{tab:notation}
	\begin{tabular}{p{0.15\textwidth}p{0.75\textwidth}}
	\hline
	Symbol & Description \\
	\hline
	$N$		&   Number of components 	\\
	$T$		&   Number of templates	\\
	$B$		&   Number of filter bands	\\
	$z_\alpha$		&  Redshift of component $\alpha$	\\
	$m_{0, \alpha}$	&  Reference band magnitude of component $\alpha$	\\
	$t_\alpha$		&  Template index of component $\alpha$	\\
	$\{z\}$		& Set of redshifts of each component	\\	
	$\{m_0\}$		& Set of reference band magnitudes of each component	\\
	$\{t\}$		& Set of template indices of each component	\\
	$b$		&   Index over filter bands	\\
	$b_0$	&   Index of reference band filter	\\
	$\hat F_0$		&   Observed flux in reference band	\\
	$\hat{\vct{F}} $		&   Vector of observed fluxes, excluding the reference band 	\\
	$\sigma_{0}$		&   Error on the reference band flux\\
	$\sigma_{b}$		&   Error on the flux in band $b$\\
	$F^{(1)}_{ t, b } \big(z, m_0 \big)$		&   Model flux for a single component in band $b$, at redshift $z$, with reference band magnitude $m_0$ and templates $t$\\
	$F^{(N)}_{ \{t\}, b } \big(\{z\}, \{m_0\} \big)$ &   Model flux for $N$-component blended source in band $b$, at redshifts $\{z\}$, with reference band magnitudes $\{m_0\}$ and templates $\{t\}$\\
	$\chi$		&   Set of cosmological parameters $\Omega_{\textrm{m}}$, $\Omega_\Lambda$ and  $H_0$	\\
	$\xi^{(N)}_\chi\big(\{z\}\big)$	&   Combination of up to $N$-point correlation functions describing the extra probability of $N$ galaxies jointly sitting at redshifts $\{z\}$ due to clustering \\
	\hline
	\end{tabular}
\end{table*}

This paper is organised as follows. In section~\ref{sec:formalism}, we describe our formalism for estimating redshifts as a parameter inference problem, describing its application to partially blended systems in section~\ref{sec:part-blend}. In section \ref{inference}, we discuss our inference methods, detailing how we use model comparison to identify blended objects in section~\ref{sec:model-select}. In section~\ref{sec:sim-results}, we test our method on simulated observations. Section~\ref{sec:gama-data} describes a test of our method on the Galaxy And Mass Assembly survey~\citep[GAMA, ][]{gamaData} blended sources catalogue~\citep{gamaBlends}, for which spectroscopic redshifts are available. We conclude in section~\ref{sec:conclusions}.

\section{Blended photo-z formalism} \label{sec:formalism}

\subsection{Flux model} \label{sec:flux}

In the same way as other template-based photometric redshift methods, we assume that each observed component is well represented by one of a set of $T$ templates. Each template $t$ is defined by its rest-frame spectral flux density $F_{t}(\lambda_{\textrm{em}})$ as a function of the emitted wavelength $\lambda_{\textrm{em}}$. This template is redshifted and observed through a broadband filter $b$, the response of which is denoted $W_b(\lambda_{\textrm{obs}})$ as a function of observed wavelength $\lambda_{\textrm{obs}}$.

The flux of template $t$, at redshift $z$ and observed in band $b$ is then given by 
\begin{equation}
T_{t,b}(z) = \frac{1}{c g^{\rm{AB}} C_b} \int_0^\infty F_{t} \left(\frac{\lambda}{1+z} \right) W_b(\lambda) \lambda \textrm{d} \lambda \,,
\end{equation}
where $g^{\rm{AB}}=3631 \textrm{ Jy}$ is the zero-point of the AB-magnitude system and the normalisation $C_b \equiv \int_0^\infty \frac{W_b(\lambda)}{\lambda} \textrm{d} \lambda$. By including $g^{\rm{AB}}$, our fluxes are dimensionless throughout, and the conversion between magnitudes and fluxes defined in the way is given by $F \equiv 10^{-0.4 m}$. This template is then scaled by a normalisation $a$ so that the flux of an object modelled with template $t$, at a redshift $z$ and observed in band $b$ is given by
\begin{equation}
F^{(1)}_{t, b} \big(z, a \big) = a T_{t, b} \big(z \big) \,.
\end{equation}
We model the flux of blended sources as a linear combination of individual component fluxes. For a blend of $N$ components, the flux observed in band $b$ is given by
\begin{equation}
F^{(N)}_{ \{t\}, b } \big(\{z\}, \{a\}\big) = \sum_{\alpha=1}^{N} a_\alpha T_{t_\alpha, b} \big(z_\alpha \big) \,,
\end{equation}
where $a_\alpha$ is the normalisation for component $\alpha$. For the reasons specified in section~\ref{sec:priors}, we sample $m_{0, \alpha}$, the apparent magnitude of each component in the reference band $b_0$ rather than this normalisation directly. The normalisation $a_\alpha$ is then defined such that the model flux in the reference band is equal to $m_{0, \alpha}$. Thus, the model flux is given by
\begin{equation}
\label{eqn:blend-flux-model}
F^{(N)}_{ \{t\}, b } \big(\{z\}, \{m_0\} \big) =  \sum_{\alpha=1}^{N} \frac{10^{-0.4 m_{0, \alpha}}}{ T_{t_\alpha, b_0} \big(z_\alpha \big)} T_{t_\alpha, b} \big(z_\alpha \big) \,.
\end{equation}

\subsection{Fully-blended posterior} \label{sec:full-blend}

For a fixed number of components, photometric redshift determination is a parameter inference problem; we wish to infer the joint posterior distribution of the redshifts and apparent magnitudes of each component given a data vector $\hat{\vct{D}}$ of $B$ broadband fluxes. This data vector is split into two parts $\hat{\vct{D}} = (\hat{\vct{F}} \, , \hat F_0)$, where $\hat F_0$ is the flux of the reference band and $\hat{\vct{F}}$ is the vector of the remaining $B-1$ fluxes. This is done since the normalisation of each component is defined in the reference band, and it is the flux of this band on which the priors are conditioned. 

\resp{Following BPZ~\citep{bpz}, we set the flux of non-detections to zero. Likewise, bands that are not observed are given a flux of zero, with the corresponding error set to an extremely large value. As discussed in section~\ref{sec:selection_function}, we assume that sources are selected using a magnitude limit on a single selection band. We therefore require that the source is detected in this band, by definition.}

We start by writing our desired posterior as a marginalisation over templates for each component. For $N$ components, we marginalise over sets of $N$ template indices $\{t\}_i = \{t_\alpha, t_\beta \dots t_{N}\}_i$. Each template index can take a value $1 \leq t \leq T$ and components may share the same template, so there are $T^{N}$ of these sets to marginalise over, giving
\begin{equation}
P\Big(\{z\}, \{m_0\}  \,\Big|\, \hat{\vct{F}}, \hat F_0, \chi, N\Big)  = \! \sum_{i=1}^{T^{N}} P\Big(\{z\}, \{t\}_i, \{m_0\} \,\Big|\, \hat{\vct{F}}, \hat F_0, \chi, N\Big) \,. \!\!\!
\end{equation}

We have emphasised that our posterior is defined for a fixed number of components by conditioning on $N$. In the general case where this number is unknown \textit{a priori}, it can be inferred from the data; this is discussed in section~\ref{sec:model-select}. We have also made the dependence on cosmological parameters, which are required for converting between distance and redshift, explicit in the above expression. These parameters are denoted by $\chi = \{\Omega_{\textrm{m}}, \Omega_\Lambda, H_0\}$ for brevity. Applying Bayes rule, the posterior becomes 
%
%
\begin{equation}
\begin{aligned}
& P\Big(\{z\}, \{m_0\}  \,\Big|\, \hat{\vct{F}}, \hat F_0, \chi, N\Big) 
\propto \\
& \indent \sum_{i=1}^{T^{N}} 
P \Big(\hat{\vct{F}}, \hat F_0 \,\Big|\, \{z\}, \{t\}_i, \{m_0\} , N \Big) 
P \Big(\{z\}, \{t\}_i, \{m_0\}  \,\Big|\, \chi, N \Big) \,. \!
\end{aligned}
\end{equation}
Since only the prior is dependent on cosmological parameters, we have removed the conditioning on $\chi$ from the likelihood. We then factorise the likelihood so that it is split in the same way as the data vector, giving
\begin{equation}
\begin{aligned}
&
P\Big(\{z\}, \{m_0\}  \,\Big|\,  \hat{\vct{F}}, \hat F_0, \chi, N\Big)
 \propto    \sum_{i=1}^{T^{N}} 
\;  P \Big(\hat{\vct{F}} \,\Big|\, \{z\}, \{t\}_i, \{m_0\} , N\Big) \;\times \\
 & \indent
P \Big(\hat F_0 \,\Big|\, \{m_0\}, N\Big) 
 P \Big(\{z\}, \{t\}_i, \{m_0\}  \,\Big|\, \chi, N \Big) \,. 
\\
\end{aligned}
\end{equation}

Since the magnitude of each component in the reference band is a sampled parameter in the posterior, our model for the reference band flux is simply the sum of these after converting from magnitudes to fluxes. As a result, the conditioning on $\{z\}$ and $ \{t\}_i$ in the reference band likelihood is unnecessary and so has been removed. We assume that the error on the observed reference band flux is normally distributed with variance $\sigma_{0}^2$. Thus, the reference band likelihood is given by
\begin{equation}
\label{eqn:ref-band-like}
P \Big(\hat F_0 \,\Big|\, \{m_0\}, N \Big) = 
\frac{1}{\sqrt{2 \pi \sigma_0^2}}
\textrm{exp} \left[ \frac{ - \left(\hat F_0 - \sum_{\alpha=1}^{N} 10^{-0.4 m_{0, \alpha}} \right)^2 }{ 2 \sigma_{0}^2} \right] \,,
\end{equation}
where $m_{0, \alpha}$ is the sampled reference band magnitude for component $\alpha$. Similarly, we use an uncorrelated multivariate Gaussian likelihood for $\hat{\vct{F}}$, 
\begin{equation}
\begin{aligned}
&
P \Big(\hat{\vct{F}} \,\Big|\, \{z\}, \{t\}_i, \{m_0\}, N  \Big)
=
 \\  & \indent  
\prod_{b=1}^{B}
\frac{1}{\sqrt{2 \pi \sigma_b^2}}
\textrm{exp} \left[ -  \frac{  \left(\hat F_b - F^{(N)}_{ \{t\}_i, b } \big(\{z\}, \{m_0\} \big) \right)^2 }{ 2 \sigma_{b}^2} \right] \,,
\end{aligned}
\end{equation}
where $F^{(N)}_{ \{t\}, b } \big(\{z\}, \{m_0\} \big)$ is the model flux is specified in equation~\ref{eqn:blend-flux-model} and $\sigma_{b}^2$ is the variance on the observed flux in band $b$. 

\subsection{Separating the joint prior}
\label{sec:priors}

We now develop the prior so that it can be written in terms of individual components. We start by separating the joint prior into a product over priors on redshift, template and magnitude. Removing unnecessary conditioning, the joint prior becomes
\begin{equation}
\label{eqn:split-prior}
\begin{aligned}
&
P \Big(\{z\}, \{t\}, \{m_0\}  \,\Big|\, \chi, N \Big) = 
 P \Big(\{z\} \,\Big|\, \{t\}, \{m_0\}, \chi, N \Big) \;\times \\
& \indent
P \Big(\{t\} \,\Big|\,  \{m_0\}, N \Big)
P \Big( \{m_0\} \,\Big|\, N  \Big) \,.
\end{aligned}
\end{equation}

This splitting up of the joint prior is similar to the approach of \cite{bpz}. There are two important differences, however. Firstly, we include a prior on the apparent magnitude of each component. This differs from the approach of \cite{bpz} who considers the magnitude on which the redshift and template priors are conditioned to be exactly the observed reference band magnitude. The uncertainty in the scaling of the template is then represented by marginalising over a normalisation factor with an assumed flat prior. However, while this normalisation is not defined as such, it is acting to set the apparent magnitude of the source in the reference band. This magnitude is a quantity about which prior information is known.

The prior information on the apparent magnitude of components is particularly important in the blended case, as we need to consider more than just the overall magnitude of the source. The individual magnitudes of each component are necessary for scaling the model fluxes when predicting the model flux $F^{(N)}_{ \{t\}, b } \big(\{z\}, \{m_0\} \big)$. In addition, motivated by existing galaxy observations and following \cite{bpz}, our redshift and template priors for each component are magnitude-dependent. The individual component magnitudes are not directly observed in the blended case, and must therefore be considered as random variables in our model. 

An alternative to sampling the magnitudes directly would be to make the fraction each component contributes the total flux a model parameter. However, the combination of intrinsic magnitude distributions and survey-specific selection effects would give the distribution of this fraction a highly complicated shape. Instead, including a prior on the magnitude of each component allows these effects to be easily accounted for.

The other important difference in the blended case is that each term in equation~\ref{eqn:split-prior} is a joint prior over all components in the source. The redshift, type and magnitude properties of individual galaxies are much more well studied than those of blended sources. To make use of this information, we write these joint priors in terms of priors on the individual components.

Firstly, we assume that the template priors for each component are independent, i.e., galaxy types are not correlated. This allows us to split the template prior as
\begin{equation}
 P \Big(\{t\} \,\Big|\,  \{m_0\}, N \Big) = \prod_{\alpha=1}^{N} P \Big(t_\alpha \,\Big|\,  m_{0, \alpha} \Big) \,.
\end{equation}

We also make the assumption that the redshift of each component depends only on its own type, not the types of other components. The redshifts of each component cannot be assumed to be independent however, as galaxies are distributed in a correlated way. The additional probability of finding $N$ galaxies within a separation $r$ over a random Poisson process is described by galaxy correlation functions of up to order $N$~\citep{xiDefinition}. We denote the combination of correlation functions describing this extra correlation as $\xi^{(N)}_\chi\big(\{z\}\big)$, i.e., the excess probability for two galaxies is given by 
\begin{equation}
 1 + \xi^{(2)}_\chi (z_\alpha, z_\beta)  \equiv  1 + \xi(r_{\alpha \beta})  \,,
\end{equation}
where the separation $r_{\alpha \beta} \equiv |\vec r_\alpha - \vec r_\beta|$ is the comoving distance between components $\alpha$ and $\beta$. In the two-component case, only the two point correlation function $\xi(r)$ is necessary. However for three galaxies, higher order correlation functions are needed, i.e.,
\begin{equation}
\begin{aligned}
&
 1 + \xi^{(3)}_\chi(z_\alpha, z_\beta, z_\gamma)  \equiv  1 + \xi(r_{\alpha \beta}) + \xi(r_{\beta \gamma}) + \xi(r_{\alpha \gamma}) \\
& \indent
+ \zeta(r_{\alpha \beta}, r_{\beta \gamma}, r_{\alpha \gamma})  \,,
\end{aligned}
\end{equation}
where $\zeta(r_{\alpha \beta}, r_{\beta \gamma}, r_{\alpha \gamma})$ is the connected three-point galaxy correlation function.

The excess probability term  $\xi^{(N)}_\chi \big(\{z\}\big)$ is defined in the posterior as a function of the component redshifts $\{z\}$, though the galaxy correlation function $\xi$ (and higher order correlations) are defined in terms of comoving separation $r$. We therefore need to convert between the redshifts of each component and the comoving distance separating them. The line of sight comoving distance as a function of redshift is given by~\citep[e.g.,][]{hoggDistance}
\begin{equation}
r(z) = \frac{c}{H_0} \int_{0}^{z} \frac{\textrm{d} z'}{E(z')} \,,
\end{equation}
where, neglecting radiation density and rewriting $\Omega \equiv \Omega_{\textrm{m}} + \Omega_{\Lambda}$, 
\begin{equation}
E(z) = \sqrt{\Omega_{\textrm{m}} (1+z)^3 + (1 - \Omega) (1+z)^2 + \Omega_\Lambda} \,.
\end{equation}
We assume a flat Planck\footnote{We use the TT + lowP + lensing + ext values from Table 4.}~\citep{planck} cosmology throughout; $\Omega_{\textrm{m}} = 0.3065$, $\Omega_\Lambda = 0.6935$ and $H_0 = 67.9 \, \textrm{km} \, \textrm{s}^{-1} \, \textrm{Mpc}^{-1}$.

However, the comoving distance separating components will depend not only on their redshifts, but also on their angular separation on the sky. As a result, we derive an effective correlation function $\xi_{\rm{eff}}$ that takes this angular dependence into account.

\begin{figure}
	\includegraphics[width=\columnwidth]{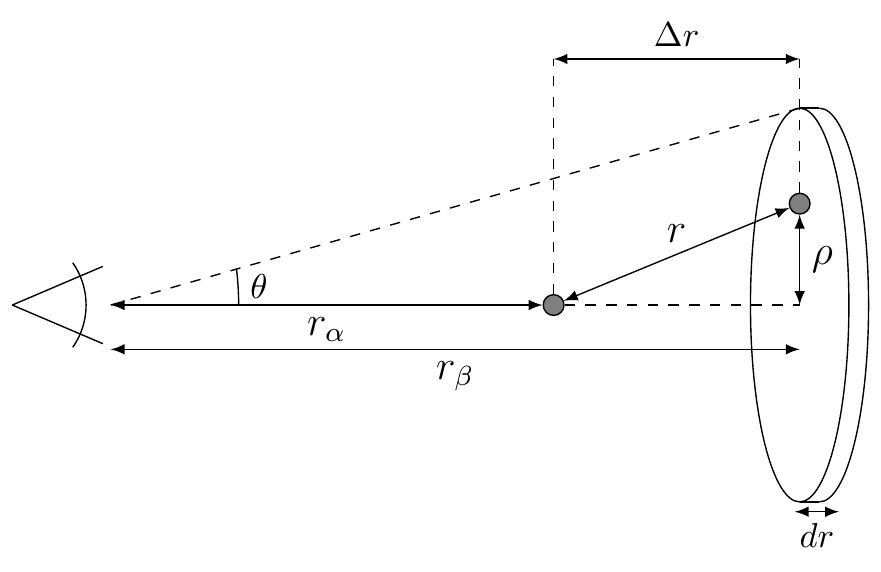}
	\caption{Diagram showing the setup of the $\xi_{\rm{eff}}$ calculation. We assume that two galaxies, represented by grey circles, will be blended if their angular separation is within $\theta$. Given that these two galaxies are blended, the galaxy at a comoving distance $r_\beta$ will lie within the disc.}
	\label{fig:xiEffDiagram}
\end{figure}

Consider the case of a two-component blend, as shown in Fig.~\ref{fig:xiEffDiagram}. The two components are at comoving distances $r_\alpha$ and $r_\beta$ from the observer, with separation $\Delta r \equiv r_\beta - r_\alpha$. From the definition of the correlation function, we can write the ratio of the expected number of galaxies in a region with clustering $N^{\xi}$ and that without $N^{0}$ as
\begin{equation}
1 + \xi_{\rm{eff}} = \frac{N^{\xi}}{N^{0}} \,.
\end{equation}

Given that these components are blended, there is some maximum angular separation $\theta$ between them; we assume this to be small. We therefore compare the expected number of galaxies in a disc of width $dr$ and radius $\rho_{\rm{max}} = r_\beta \theta$. The expected number without clustering is given by
\begin{equation}
N_{\rm{disc}}^0 = \bar n \pi r_\beta^2 \theta^2 \,.
\end{equation}

To find the expected number with clustering, we integrate over the disc using the volume element of an annulus with radius $\rho$, i.e.,
\begin{equation}
N_{\rm{disc}}^\xi = \int_{\rho=0}^{r_\beta \theta} \bar n [1 + \xi(r)] \, 2 \pi \rho \, d \rho \, d r \,.
\end{equation}
Thus, writing $r = \sqrt{\Delta r^2 + \rho^2}$, the ratio becomes
\begin{equation}
\label{eqn:xi_eff_integral}
1 + \xi_{\rm{eff}} = \frac{2}{r_\beta^2 \theta^2}  \int_{\rho=0}^{r_\beta \theta} \left[1 + \xi\left(\sqrt{\Delta r^2 + \rho^2} \right) \right] \, \rho \, d \rho \,.
\end{equation}

As described below, the effect of clustering is small. As a result, we adopt a simple power law for the two point correlation function,
\begin{equation}
\xi(r) \propto \left( \frac{r}{r_0} \right)^{-\gamma} \,.
\end{equation}
Inserting this into equation~\ref{eqn:xi_eff_integral} and integrating, the effective correlation function is given by
\begin{equation}
\xi_{\rm{eff}} (r_\alpha, r_\beta) = \frac{r_0^2}{\left(1 - \frac{\gamma}{2} \right) r_\beta^2 \theta^2} \left[ \left( \frac{\Delta r^2 + r_\beta^2 \theta^2}{r_0^2} \right)^{1 - \frac{\gamma}{2}} - \left( \frac{\Delta r^2}{r_0^2} \right)^{1 - \frac{\gamma}{2}}\right] \,.
\end{equation}

\resp{The effect of the strength of clustering evolving with redshift can be included in this formalism by allowing the parameters $r_0$ and $\gamma$ to vary with redshift~\citep[e.g.][]{xiEvolution}. We test the effect of this on the redshift inference by using a toy model where $\gamma=1.92$ is kept constant, while $r_0$ linearly varies between $r_0 = 5 \, \textrm{Mpc} \, h^{-1}$ at redshift $z=2$ and $r_0 = 6 \, \textrm{Mpc} \, h^{-1}$ at redshift $z=0.5$, with a linear extrapolation outside of this range. Since the value of $\xi_{\rm{eff}}$ is non-negligible only when $z_\alpha \approx z_\beta$, this interpolation of $r_0$ is evaluated using $z_\alpha$ only.}

\begin{figure}
	\includegraphics[width=\columnwidth]{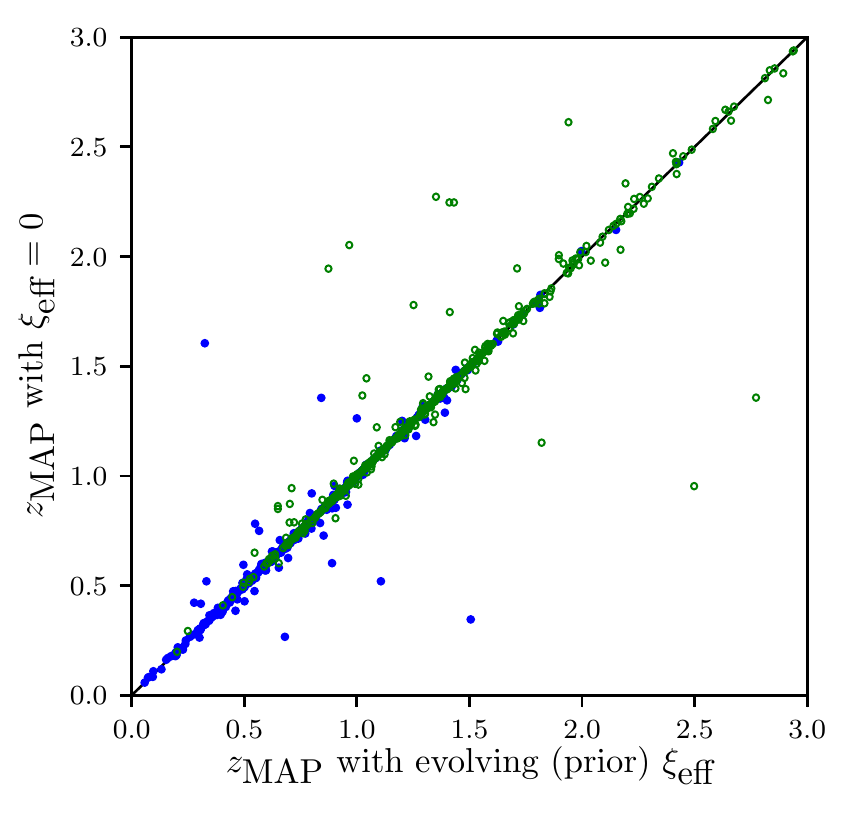}
	\caption{\resp{Comparison of the maximum \textit{a posteriori} point estimates including the effective correlation function and neglecting it, for sources simulated from a prior that includes it. The lower redshift components $z_\alpha$ are plotted with closed blue markers, and $z_\beta$ are plotted with open green markers. Most sources show negligible differences, while sources that show large differences are multimodal. In these sources, small differences in the posterior result in point estimates moving between modes of slightly different heights, illustrating a limitation of point estimates.} }
	\label{fig:xiVsNoXi}
\end{figure}

{We then} simulated two-component blends from a prior with $\xi_{\rm{eff}}$ included as described in section~\ref{sec:sim-results}. Results assuming $\xi_{\rm{eff}}=0$ showed negligible differences from those where the effect was included. \resp{At the population level, the RMS scatter defined in equation~\ref{eqn:rms_scatter} changed by $0.205\%$ between results including and excluding the correlation function. There were also negligible changes to the results at the individual source level. A comparison of maximum \textit{a posteriori} results in each case are shown in Fig.~\ref{fig:xiVsNoXi}. The vast majority of sources show negligible differences, and visually inspecting the posteriors with larger changes shows these are highly multimodal, with modes of comparable heights. In these cases, small differences in the posteriors result in larger differences in point estimates as the maximum \textit{a posteriori} value moves between modes. This is a limitation of point estimates, and can be mitigated by using the full information content of the posterior distributions, which do not vary strongly.}
	
\begin{figure}
	\includegraphics[width=\columnwidth]{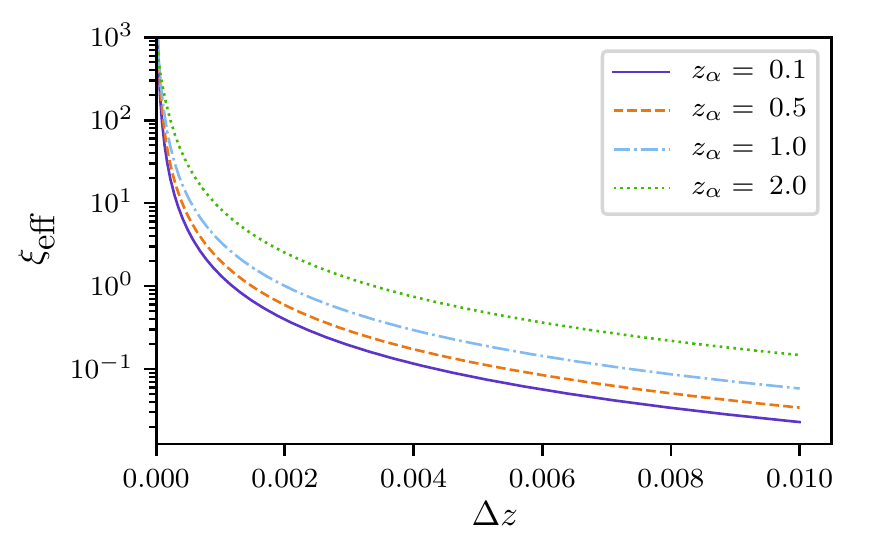}
	\caption{Plot of the effective correlation function $\xi_{\rm{eff}}$ vs $\Delta z \equiv z_\beta - z_\alpha$ for various $z_\alpha$ {used for the results throughout}.}
	\label{fig:xiEffPlot}
\end{figure}

\resp{Due to the small effect, our results throughout include a simple non-evolving correlation function with $r_0=5 \, \textrm{Mpc} \, h^{-1}$ and $\gamma=1.77$~\citep{xiDefinition}. A plot of this is given in Fig.~\ref{fig:xiEffPlot}.}

{Inserting the correlation function} allows us to write the joint redshift prior {separated by component} as
\begin{equation}
\begin{aligned}
P \Big(\{z\} \,\Big|\, \{t\}, \{m_0\}, \chi, N \Big) =  \left[ 1 + \xi^{(N)}_\chi \big(\{z\} \big) \right] \prod_{\alpha=1}^{N} P \Big(z_\alpha \,\Big|\, t_\alpha, m_{0, \alpha} \Big) \,.
\end{aligned}
\end{equation}

We separate the joint magnitude prior by assuming that the only correlation between the component magnitudes is from the effect of a selection function $S\left( \{m_0\}\right)$ applied to the total magnitude, as discussed in section~\ref{sec:selection_function}. The magnitude prior can then be written as
\begin{equation}
\label{eqn:magnitude-prior}
P \Big( \{m_0\} \,\Big|\, N  \Big) = S\Big( \{m_0\}\Big) \prod_{\alpha=1}^{N} P \Big( m_{0, \alpha} \Big) \,,
\end{equation}

\begin{figure}
	\includegraphics[width=\columnwidth]{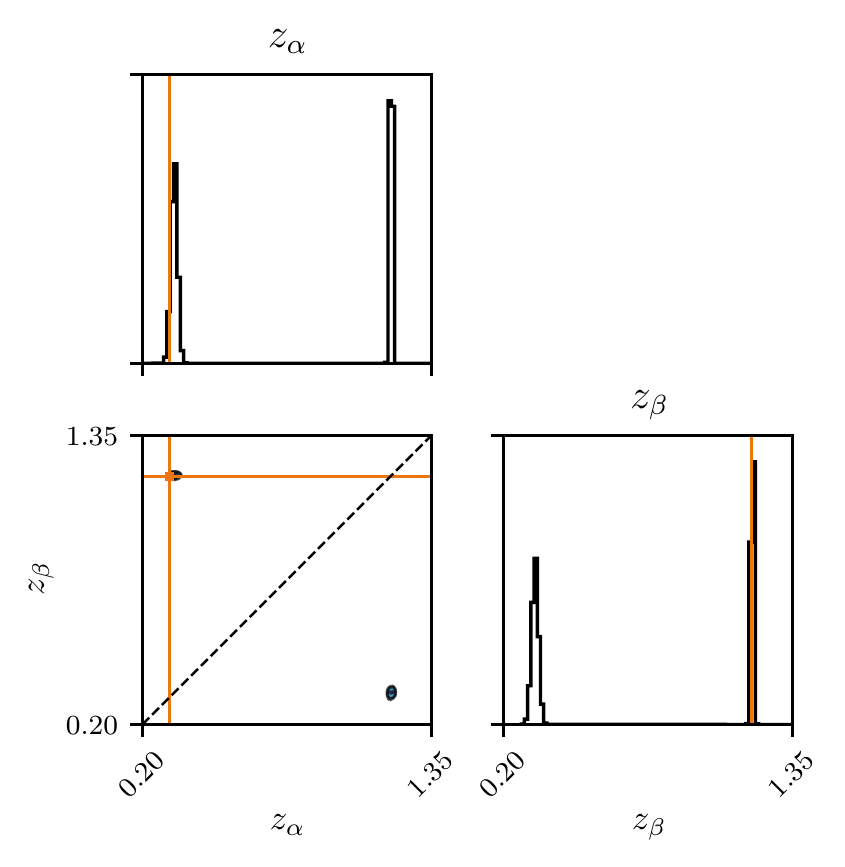}
	\caption{By not imposing a sorting condition, the components in a source are exchangeable. This is demonstrated here for a simple two-component blend with redshifts $z_\alpha=0.31$, $z_\beta=1.19$ as indicated by the orange lines. As a result of the exchangeability, the 2D marginal redshift distribution is symmetric about the dashed black line, and each 1D posterior contains a distinct peak for each component.}
	\label{fig:redshift-exchange}
\end{figure}

Finally, we impose a sorting condition. Without this, the components would be exchangeable, i.e., swapping the component labels $\alpha, \beta \dots$ would have no effect on the prediction of the model. As a result, the marginalised posterior for the redshift of a single component would contain contributions from every component in the source, as demonstrated in Fig.~\ref{fig:redshift-exchange}.

\label{sec:redshift-degeneracy}
Imposing a sorting condition on either the magnitudes or the redshifts would have the same effect of breaking the exchangeability of the components. In our tests, sorting by redshift produced posteriors that recovered the true redshift more successfully. However, in high redshift samples, there is an intrinsic colour degeneracy that can occasionally cause problems with a redshift sorting condition. 

The Lyman break and Balmer break are absorption features occurring at $912$\AA\, and $3650$\AA\, respectively. If photometry over a sufficiently wide wavelength range is not available, a Lyman break at high redshift can be confused with a Balmer break at low redshift~\citep[e.g.,][]{lymanBalmer}. If the sample is deep enough that these high redshift solutions are not unlikely \textit{a priori}, this can cause bimodal posteriors and contribute to catastrophic outliers~\citep{degeneracyDepth}.

Consider the case of a two-component blend where the redshift of one component is well constrained but the other has a bimodal posterior. If the well constrained redshift happens to lie between these two modes, it will appear in the 1D marginal distributions of each component redshift, as whether it is the lower or higher redshift object depends on which of the two degenerate peaks is being sampled. In this case, sorting by magnitudes would result in a posterior more representative of the underlying system, where the redshift of one component is well constrained while the other has two well separated modes. We did not find this to be a problem in our tests however, and so apply redshift sorting throughout.

The sorting condition $\Lambda_\alpha$ is imposed by introducing Heaviside step functions $\Theta$ into the product over components, and is defined as
\begin{equation}
\begin{aligned}
\Lambda_\alpha &= 1 \quad \textrm{for} \quad \alpha = 1\\
&=  \Theta(q_{\alpha-1} - q_{\alpha}) \quad \textrm{otherwise,}\\
\end{aligned}
\end{equation}
where $q$ is either $z$ or $m_0$ depending on whether redshift or magnitude sorting is used. In summary, the posterior for the fully-blended case is given by
\begin{equation}
\label{eqn:posterior}
\begin{aligned}
&
P\Big(\{z\}, \{m_0\}  \,\Big|\, \hat{\vct{F}}, \hat F_0, \chi, N\Big)
\propto  \sum_{i=1}^{T^{N}} 
\; P \Big(\hat{\vct{F}} \,\Big|\, \{z\}, \{t\}_i, \{m_0\}, N  \Big) \;\times \\
& \indent 
P \Big(\hat F_0 \,\Big|\, \{m_0\} \Big) 
 \left[ 1 + \xi^{(N)}_\chi \big(\{z\} \big) \right] S\Big( \{m_0\}\Big) \;\times
 \\  & \indent 
\prod_{\alpha=1}^{N} \Lambda_\alpha 
 P \Big(z_\alpha \,\Big|\, t_\alpha, m_{0, \alpha} \Big) 
 P \Big(t_\alpha \,\Big|\,  m_{0, \alpha} \Big)
P \Big( m_{0, \alpha} \Big) \,.
\\
\end{aligned}
\end{equation}

\subsection{Accounting for selection effects}
\label{sec:selection_function}

When considering the total apparent magnitude of a source, we must account for the selection effect of the survey observing it. Galaxy surveys typically select sources by imposing cuts on the apparent magnitude they observe $m < m_{\rm{lim}}$ since they cannot observe arbitrarily faint sources. As we are sampling \textit{intrinsic} magnitudes rather than \textit{observed} magnitudes, these selection effects do not impose a hard cut in our magnitude prior. 

Consider a source with an intrinsic apparent magnitude exactly equal to the survey magnitude limit. Assuming a normal distribution for the observational error, the probability of observing this source is ${1}/{2}$, since its observed apparent magnitude is equally likely to have been scattered above and below the magnitude cut. However, since objects in the sample have been detected by definition, we know the source must have been scattered brighter, effectively breaking the symmetry of the error distribution. As a result, intrinsic apparent magnitudes around the magnitude limit are less probable and should be downweighted.

To account for this, we follow the approach described in \cite{selection} for including a selection effect. A discrete variable $D$ representing the fact that an object was detected is introduced, and each term in the posterior is conditioned on it. 
We assume that our selection effect is imposed on \resp{a single selection band. Without loss of generality, we derive the effect by assuming that the selection band is the reference band $b_0$} and so only the reference band likelihood is affected.
Conditioning on $D$, the likelihood can be written using Bayes rule as
\begin{equation}
\label{eqn:selection-bayes}
P \Big(\hat F_0 \,\Big|\, \{m_0\}, N, D \Big)  =  \frac{ P \Big(D \,\Big|\, \hat F_0,  \{m_0\}, N \Big) P \Big(\hat F_0 \,\Big|\, \{m_0\}, N \Big) } { \int_0^\infty P \Big(D \,\Big|\, \hat F_0,  \{m_0\}, N \Big) P \Big(\hat F_0 \,\Big|\, \{m_0\}, N \Big) \textrm{d} \hat F_0} \,.
\end{equation}

The numerator of equation~\ref{eqn:selection-bayes} is equal to the likelihood defined in equation~\ref{eqn:ref-band-like} since the probability of detection for an object that we know has been observed is $P \Big(D \,\Big|\, \hat F_0,  \{m_0\}, N, \Big) = 1$. After integrating over $\hat F_0$, the denominator depends only on $\{m_0\}$ and represents the effect of the magnitude selection. We therefore choose to write this term as part of the joint magnitude prior, defining the selection effect
\begin{equation}
S\Big(\{m_0\}\Big) = \int_0^\infty P \Big(D \,\Big|\, \hat F_0,  \{m_0\}, N, \Big) P \Big(\hat F_0 \,\Big|\, \{m_0\}, N \Big) \textrm{d} \hat F_0 
\end{equation}
that appears in the posterior in equation~\ref{eqn:posterior}. The selection is a hard cut based on the observed flux, and so 
\begin{equation}
\begin{aligned}
P \Big(D \,\Big|\, \hat F_0,  \{m_0\}, N, \Big) &= 1 \quad \textrm{for} \quad \hat F_0 > 10^{-0.4 m_{\textrm{lim}}}\\
&= 0 \quad \textrm{otherwise.} 
\end{aligned}
\end{equation}
Thus, the integral becomes
\begin{equation}
\label{eqn:selection-integral}
S\Big(\{m_0\}\Big) = \int_{ 10^{-0.4 m_{\textrm{lim}} }}^\infty P \Big(\hat F_0 \,\Big|\, \{m_0\}, N \Big) \textrm{d} \hat F_0 \,.
\end{equation}

Since the reference band likelihood is assumed Gaussian, this can be written in terms of the normal cumulative distribution function as $S\left(\{m_0\}\right) = 1 - \Phi(\hat F_0)$, where $\Phi$ is defined for a Gaussian distribution with mean $\mu$ and standard deviation $\sigma$ to be
\begin{equation}
\Phi(x) = \frac{1}{2} \left[ 1 + \textrm{erf} \left( \frac{x - \mu}{\sigma \sqrt{2}} \right) \right] \,.
\end{equation}

Inserting this into equation~\ref{eqn:selection-integral}, the effect of the magnitude selection can be written as
\begin{equation}
S\Big(\{m_0\}\Big) = \frac{1}{2} - \frac{1}{2} \, \textrm{erf} \left( \frac{ 10^{-0.4 m_{\rm{lim}}} - \sum_{\alpha=1}^{N} 10^{-0.4{m_{0, \alpha}}}}{\sigma_{0} \sqrt{2}} \right) \,.
\end{equation}

\resp{By replacing the reference-band flux $\sum_{\alpha=1}^{N} 10^{-0.4{m_{0, \alpha}}}$ with the model flux $F^{(N)}_{ \{t\}, b } \big(\{z\}, \{m_0\} \big)$, this selection can performed on any band. The selection function would then also be dependent on the redshifts and templates, i.e., $S\Big(\{z\}, \{t\}, \{m_0\}\Big)$. This choice of selection band is included in the implementation described in section~\ref{sec:blendz}.}

\begin{figure}
	\includegraphics[width=\columnwidth]{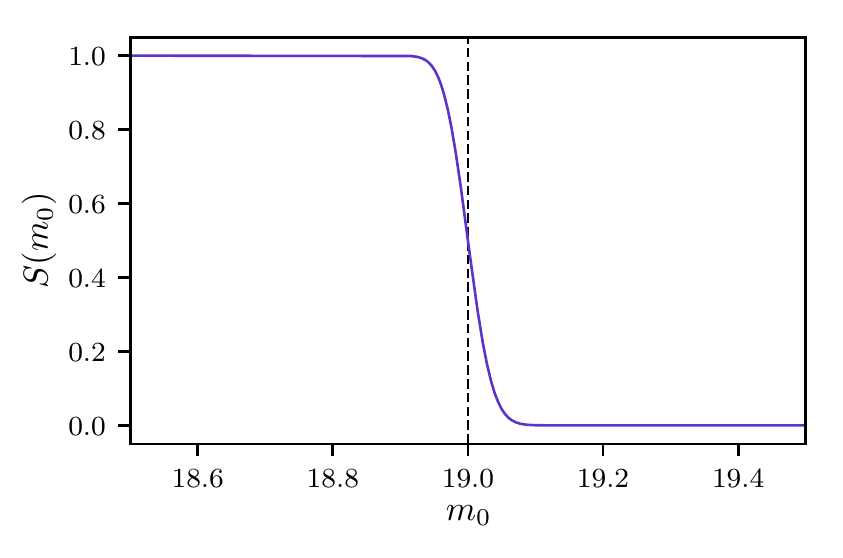}
	\caption{Plot of the selection function for a typical source from the GAMA blended sources catalogue used in section~\ref{sec:gama-data}. The dashed line shows the magnitude limit for this source $m_{\textrm{lim}} < 19$.}
	\label{fig:selectionFunction}
\end{figure}

A plot of this selection function for a galaxy from the GAMA blended sources catalogue, described in section~\ref{sec:gama-data}, is shown in Fig.~\ref{fig:selectionFunction}.

\subsection{Specifying the priors}
\label{sec:specify-priors}

Like all Bayesian methods, the choice of priors should be problem dependent. For ease of comparison, we use the parametric forms given by \cite{bpz} with an additional magnitude prior. However, we stress that this choice is not a necessary one for our method and any joint $P \Big(z, t, m_0\Big)$ prior may be used.

The \cite{bpz} template and redshift priors are given by
\begin{equation}
 P \Big(t \,\Big|\,  m_{0} \Big) = f_t \textrm{e}^{- k_t (m_0 - m_{\textrm{min}})} 
\end{equation}
and 
\begin{equation}
 P \Big(z \,\Big|\, t, m_{0} \Big) \propto z^{\alpha_t} \textrm{exp} \left\{ - \left[ \frac{z}{ z_{0, t} + k_{\textrm{m},t}(m_0 - m_{\textrm{min}}) } \right]^{\alpha_t} \right\}
\end{equation}
respectively, where $m_{\textrm{min}}$ is the bright-end magnitude cut as described below. The parameters $\alpha_t$, $z_{0,t}$, $k_{\textrm{m},t}$, $f_t$ and $k_t$ are set separately for early, late and irregular template types. Their values are found using the procedure discussed in section~\ref{sec:calibrate-priors} and are listed in Table~\ref{tab:prior-params}.

We use a magnitude prior given by
\begin{equation}
 P \Big(m_0 \Big) \propto 10^{ \, \phi m_0} \,.
\end{equation}
The value $\phi=0.6$ gives the expression for the expected galaxy number counts in a homogeneous, Euclidean universe \citep{numberCounts}, though we leave $\phi$ free to also be found using the procedure discussed in section~\ref{sec:calibrate-priors}. This was found to be $\phi=0.705$, though the difference in results compared to fixing $\phi=0.6$ was negligible.

Since the selection effect applies to the total source flux, individual components may be fainter than the survey magnitude limit, and so unobservable outside of a blended source. As a result, an analytic magnitude prior is required to describe the distribution of component magnitudes so that it can be used at faint magnitudes, where observations of individual components are unavailable.

For the reasons discussed in section~\ref{sec:multinest}, we also apply a hard minimum and maximum cut to each component redshift $z_\alpha$ and component magnitude $m_{0, \alpha}$. This cut has little effect on the redshift priors which already go towards zero at large redshift; the same is true of the magnitude prior at bright magnitudes. 
The faint-end of the magnitude prior of the brightest component is also already forced towards zero by the selection function. This is because in an $N$-component blend, the flux of the brightest component must be at least $1/N$ that of the total source flux, by definition. 
The magnitudes of the other components are not constrained in this way however, and so this cut represents a sharp boundary in the prior. 

In our tests, the results of the redshift estimation were not strongly dependent on the position of this faint-end  magnitude cut $m_{\textrm{max}}$. However, the evidence calculation described in section~\ref{sec:model-select} \textit{is} dependent on its position, as changing the position of the cut alters the prior volume integrated over in equation~\ref{eqn:evidence-integral}. As a result, the position of this cut must be decided; it defines the limit where a galaxy is considered to contribute to a blend, and is therefore problem-dependent.

In principle, one could consider a galaxy to be blended if another arbitrarily faint galaxy lies along the same line of sight. In practice however, observations have limited precision, and the flux of an extremely dim galaxy cannot be detected. In other words, a sufficiently dim galaxy should no longer be considered a blended component, but rather a contribution to the noise. 

In practice, a simple method to set this cut is to fix it for the entire sample. However, the argument above suggests that this cut should be dependent on the noise of the observation, i.e., that $m_{\textrm{max}}$ should be set to the faintest magnitude that would have an observable effect. Fixing $m_{\textrm{max}}$ is effectively an assumption that the sample has sufficiently homogeneous noise properties that the change in this faintest magnitude is negligible. For a sample where this is not the case, the magnitude cut can be set as an $n \sigma_0$ flux deviation, i.e.,
\begin{equation}
\label{eqn:mag-max-sigma}
m_{\textrm{max}} = - 2.5 \log_{10} (n \sigma_0) \,,
\end{equation}
where $\sigma_0$ is the error on the reference band flux which varies for each source. In the tests in section~\ref{sec:gama-data}, we test both of these methods of setting $m_\textrm{max}$.

\subsection{Calibrating the priors using spectroscopic information}
\label{sec:calibrate-priors}

The joint prior is conditioned on a set of parameters $\theta$, i.e., $P \Big(z, t, m_0 \,\Big|\, \theta \Big)$, the posterior distribution of which we wish to infer.
We can use spectroscopic information of a sample of galaxies from the population of interest to calibrate the above priors as suggested by \cite{bpz}.

We assume here that this calibration is done with unblended galaxies, though this procedure can be extended to include blended galaxies too, provided that the number of components $N$ is known \textit{a priori}. In that case, the reference band magnitudes of each component would need to be included as a parameter in this model, and either sampled along with $\theta$ or marginalised out of the posterior analytically. 

We consider a sample of $G$ galaxies with photometry and spectroscopic redshifts $\hat z_{s}$. These redshifts are assumed to be exact, i.e., we neglect the error on $\hat z_{s}$. The set notation here now runs over each independently observed galaxy, not the blended components as before.

We start by writing this posterior as a marginalisation over the photometric redshift model parameters for each galaxy and applying Bayes rule. Since the likelihood is independent of the prior parameters, we condition on $\theta$ in the prior only, giving
\begin{equation}
\begin{aligned}
&\, P \Big(\theta \,\Big|\, \{\hat z_s\}, \{ \hat{\vct{F}}\}, \{\hat F_0\} \Big) \propto
 \int \textrm{d}^{G} \{ z\} \int \textrm{d}^{G} \{ m_0\} \;\times \\
& \indent 
\sum_{i=1}^{T^{G}}
P \Big( \{\hat z_s\}, \{ \hat{\vct{F}} \}, \{\hat F_0\} \,\Big|\, \{z\}, \{t\}_i, \{m_0\}  \Big) 
P \Big(\theta, \{z\}, \{t\}_i, \{m_0\}  \Big) \,. \!\!\!\!\!\!
\end{aligned}
\end{equation}
We apply product rule to separate the joint prior and remove other unnecessary conditioning. We also assume that the galaxies in the sample are independent, and so all terms not shared across the population (i.e., $P(\theta)$) can be written as a product over galaxies. The posterior then becomes
\begin{equation}
\begin{aligned}
 & P \Big(\theta \,\Big|\, \{\hat z_s\}, \{ \hat{\vct{F}}\}, \{\hat F_0\} \Big)
\propto P \Big(\theta \Big) \prod_{g=1}^{G} 
\int \textrm{d} z_g \int \textrm{d} m_{0, g} \;\times \\
& \indent 
\sum_{i_g =1}^{T}
P \Big( \hat F_{0, g} \,\Big|\, m_{0, g}  \Big)
 P \Big( \hat{\vct{F}}_g  \,\Big|\, z_g, t_{i_g}, m_{0, g} \Big) \;\times \\
& \indent 
P \Big( \hat z_{s, g} \,\Big|\, z_g \Big)
P \Big( z_g, t_{i, g}, m_{0, g} \,\Big|\, \theta  \Big) \,.
\end{aligned}
\end{equation}

By assuming that the spectroscopic redshifts are exact, the redshift likelihood can be written as a delta function, i.e., $P \Big( \hat z_{s, g} \,\Big|\, z_g \Big) = \delta \Big( z_{g} - \hat z_{s, g} \Big)$. We also assume that the error on the reference band magnitude is negligible, allowing us to write $P \Big( \hat F_{0, g} \,\Big|\, m_{0, g}  \Big) = \delta \Big( m_{0, g} - \hat m_{0, g} \Big)$, where $\hat m_{0, g} = -2.5 \log_{10} \left( \hat F_{0, g} \right)$ is the reference band flux of galaxy $g$, converted to magnitudes. Replacing these likelihoods with delta functions, the marginalisation can be done analytically using the sifting property of the delta function to give
\begin{equation}
\begin{aligned}
& P \Big(\theta \,\Big|\, \{\hat z_s\}, \{ \hat{\vct{F}}\}, \{\hat F_0\} \Big)
\propto 
P \Big(\theta \Big) \; \times
\\ & \indent
\prod_{g=1}^{G}
\sum_{i_g =1}^{T}
P \Big( \hat{\vct{F}}_g  \,\Big|\, \hat z_{s, g}, t_{i_g}, \hat m_{0, g}  \Big) 
P \Big( \hat z_{s, g}, t_{i_g}, \hat m_{0, g} \,\Big|\, \theta  \Big) 
\,. 
\end{aligned}
\end{equation}

To find the prior parameters $\theta$ that maximise this posterior, we use L-BFGS-B~\citep{lbfgsb}, a local optimisation algorithm that approximates the Hessian of the objective function and optimises the parameters subject to simple box constraints; we use these constraints to ensure our parameters are positive. This method requires first-order derivatives which we approximate through a finite difference method. 

\begin{figure*}
	\includegraphics{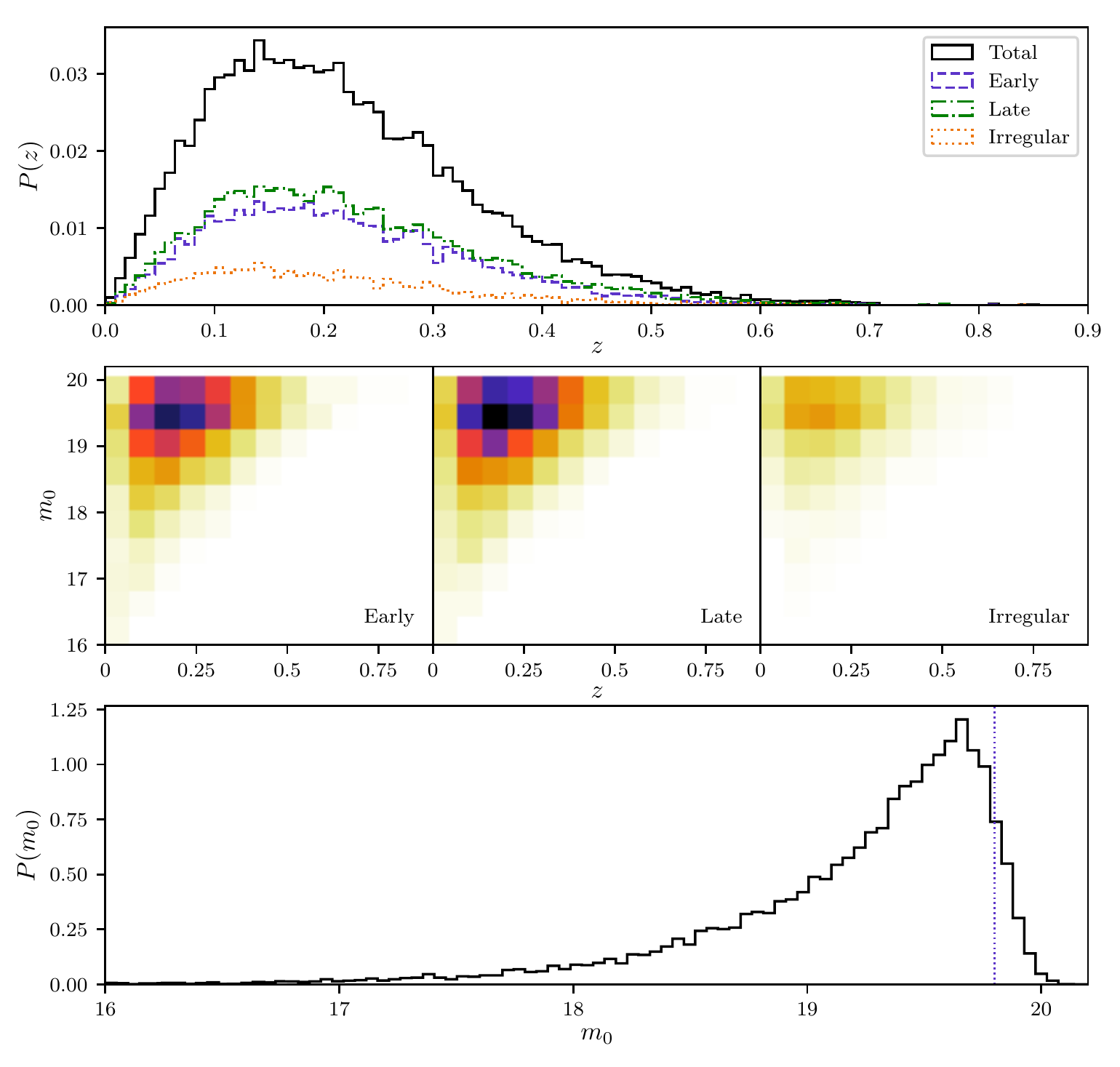}
	\caption{Plot of the prior found for the test on the GAMA blended sources catalogue after calibrating using 26782 unblended sources. The dashed line in the bottom panel shows a magnitude limit of $r < 19.8$.}
	\label{fig:gama-prior-calibration}
\end{figure*}

The result of this procedure is an estimate of the maximum \textit{a posteriori} values of the prior parameters. Throughout this paper, we use these values in the priors directly. In principle, these parameters could form part of a hierarchical model and be marginalised out as nuisance parameters. However, this would significantly increase the dimensionality of the parameter space to be sampled and, thus, the computation time required for each source. Table~\ref{tab:prior-params} lists the values of these prior parameters GAMA test described in section~\ref{sec:gama-data}. A plot of samples drawn from the resulting prior is shown in Fig.~\ref{fig:gama-prior-calibration}.

\begin{table}
	\caption{The maximum \textit{a posteriori} values of the prior parameters for the GAMA blended sources catalogue after calibrating using 26782 unblended sources. }
	\centering
	\label{tab:prior-params}
	\begin{tabular}{lllll}
		\hline
		Parameters & Early & Late & Irregular & Type-independent \\ \hline
		$\alpha_t$                & 1.59 & 1.53 & 1.30 & - \\
		$z_{0,t}$                 &  0.016 & 0.019 & 0.066 & -  \\
		$k_{m,t}$                 &   0.048 & 0.048 & 0.022 & -\\
		$k_t$                     & 0.044 & 0.024 & -  &- \\
		$f_t$                     &0.45 & 0.51 &  - & -\\ 
		$\phi$                   & -   & - & - & 0.71 \\ \hline
	\end{tabular}
\end{table}

\section{\resp{Partially-blended sources}} \label{sec:part-blend}

We can modify the {formalism} above for the case of 
\resp{
sources for which every component does not contribute to every observation. We refer to these as partially blended sources. This can be the case when combining photometry from a wide range of wavelengths, e.g., optical and far-infrared observations. This partial blending may also occur for some sources observed in both a ground-based and space-based survey, as the latter does not suffer from atmospheric seeing and so can achieve a higher spatial resolution. An example of a pair of such surveys is LSST~\citep{lsstSummary} and Euclid~\citep{euclidSummary}. Utilising resolved photometry from Euclid could improve the precision of photometric redshifts of sources that are blended in the higher signal-to-noise observations of LSST. This possibility is explored using simulated observations in section~\ref{sec:part-sim-results}.
}

To generalise the method for this case, we introduce the measurement-component mapping $\delta_{\alpha, m}$, an $N \times N_m$ matrix, where $N_m$ is the number of measurements, a generalisation of the number of bands in the fully-blended case. This measurement-component mapping acts as an indicator variable, consisting only of zeros and ones indicating whether a particular component is present in a particular measurement. 

An example of such a matrix is given below. Consider data containing $N_m = 6$ photometric measurements of $N=2$ components. The first four measurements are of individually resolved components, while the final two measurements are blended. In a typical use case, we might expect the resolved measurements of each component to share filter bands, though the model does not require this. In this example, the measurement-component mapping is given by
\begin{equation}
\delta = \begin{bmatrix}
1 & 0 & 1 & 0 & 1 & 1 \\
0 & 1 & 0 & 1 & 1 & 1 \\
\end{bmatrix} \,.
\end{equation}

We can then write the blended flux of $N$ components at a redshift $z$ in measurement $m$ as
\begin{equation}
\label{eqn:part-blend-flux-model}
F^{(N)}_{ \{t\}, m, \delta } \big(\{z\}, \{m_0\} \big) =  \sum_{\alpha=1}^{N} \delta_{\alpha, m} \frac{10^{-0.4 m_{0, \alpha}}}{ T_{t_\alpha, b_0} \big(z_\alpha \big)} T_{t_\alpha, m} \big(z_\alpha \big) \,.
\end{equation}

\resp{The only modification to the posterior of the fully-blended case needed to accommodate the partial-blending is to the sorting condition.} As described in section~\ref{sec:priors}, the purpose of this condition is to prevent the exchangeability of components However, this is not necessary in the partially blended case. Here, the components are intrinsically different as they appear individually in separate measurements and so are not exchangeable. As a result, we drop the sorting condition for the partially blended case, i.e., $\Lambda_\alpha=1$ over the entire parameter space. The posterior for the partially blended case is then given by
\begin{equation}
\label{eqn:part-posterior}
\begin{aligned}
&
P\Big(\{z\}, \{m_0\}  \,\Big|\, \hat{\vct{F}}, \hat F_0, \chi, N, \delta \Big)
\propto  \sum_{i=1}^{T^{N}} 
 P \Big(\hat{\vct{F}} \,\Big|\, \{z\}, \{t\}_i, \{m_0\}, N, \delta \Big) \;\times 
\!\!\!\!\!
\\
& \indent 
P \Big(\hat F_0 \,\Big|\, \{m_0\} \Big) 
\left[ 1 + \xi^{(N)}_\chi \big(\{z\} \big) \right] S\Big( \{m_0\}\Big) \;\times
\\  & \indent 
\prod_{\alpha=1}^{N} 
P \Big(z_\alpha \,\Big|\, t_\alpha, m_{0, \alpha} \Big) 
P \Big(t_\alpha \,\Big|\,  m_{0, \alpha} \Big)
P \Big( m_{0, \alpha} \Big) \,.
\end{aligned}
\end{equation}

\section{Inference using Nested Sampling} \label{inference}

\subsection{Determining the number of components with model comparison} \label{sec:model-select}

The posteriors in equations~\ref{eqn:posterior}~and~\ref{eqn:part-posterior} are defined for a specific number of components $N$. In general however, this number of components is not known \textit{a priori}. We therefore need a method to determine how many components are present in a source. Since our model is defined for a fixed number of components, we treat finding the number of components in a source as a model comparison problem.

Bayesian model comparison involves the calculation of the evidence $\mathcal{Z}$, an integral over the product of the prior and the likelihood~\citep[e.g.,][]{trottaBayes}. Given a data vector $\vct{d}$, a model $m$ and a set of model parameters $\{\theta\}$, the evidence is defined as
\begin{equation}
\label{eqn:evidence-integral}
\mathcal{Z} \equiv P \Big(\vct{d} \,\Big|\, m \Big) = \int P \Big(\vct{d} \,\Big|\,\{\theta\}, m \Big) P \Big(\{\theta\} \,\Big|\, m \Big) \textrm{d} \{\theta\} \,.
\end{equation}

This evidence term plays the role of the normalisation of the posterior and so is typically ignored in parameter inference problems where this normalisation is irrelevant. However, the evidence is the quantity of interest for model comparison problems. The ratio of the posterior probabilities of two models is proportional to the ratio of their evidences, a quantity known as the Bayes factor. By considering the number of components in a source as the model, we can write the relative probability of the source containing $n$ components compared to $m$ components as
\begin{equation}
\label{eqn:model-select}
\mathcal{P}_{n,m} = \frac{P \Big( N = n\,\Big|\,  \hat{\vct{F}}, \hat F_0  \Big)}{P \Big( N = m\,\Big|\,  \hat{\vct{F}}, \hat F_0  \Big)}  = 
\frac{P \Big( \hat{\vct{F}}, \hat F_0 \,\Big|\, N = n  \Big)}{P \Big(  \hat{\vct{F}}, \hat F_0 \,\Big|\,  N = m\Big)}
\frac{P \Big(N = n  \Big)}{P \Big( N = m \Big)} \,.
\end{equation}

Considering the cases of either isolated galaxies or blends of two components, the model prior ratio $P \Big(N = 2  \Big) / P \Big( N = 1 \Big)$ represents the probability that a galaxy will be blended. \cite{lsstBlendReport} estimate the number of sources observed by LSST that will be blended by convolving Hubble Space Telescope images with a Gaussian point spread function (PSF) like that of LSST. They found this number to be $45-55\%$ of the total sources observed, with $15-20\%$ of observed sources classified as catastrophic blends that would be identified as single sources by fitting a profile template to a galaxy image. \cite{lsstDensity} estimates that the rejection of blended sources will reduce the number density of LSST sources by $16\%$, though this estimate does not include the catastrophic blends of above. Studies such as these using existing high-resolution data or simulated observations can inform the blending prior ratio. Throughout this paper, we present results where this prior ratio is $P \Big(N = 2  \Big) / P \Big( N = 1 \Big) = 1$, i.e., we do not prefer either of the blended or single-component models \textit{a priori}, though this information can be trivially included.

\subsection{Nested sampling using MultiNest}
\label{sec:multinest}

Calculating the evidence directly through numerical integration presents a difficult technical problem, particularly as the number of dimensions increases. To avoid this, we use Nested sampling~\citep{nestSamp}, a Monte Carlo method for estimating the evidence while also sampling the posterior for parameter inference. Nested sampling reduces the problem of estimating the evidence to sampling a series of increasing likelihood thresholds, i.e., progressively smaller prior volumes nested within one another. Equation~\ref{eqn:evidence-integral} can then be calculated using a one-dimensional quadrature integration method over this prior volume.

The computationally difficult part of the nested sampling algorithm is sampling a new point from within the potentially complicated boundary defined by the likelihood threshold. The MultiNest sampler~\citep{multinest} does this efficiently by sampling from a collection of ellipses approximating this boundary rather than the prior itself. This collection of ellipses is formed by performing a clustering analysis on a fixed-sized set of the previous samples, known as the live points. A new sample is drawn from these ellipses, replacing the lowest likelihood point which is removed and stored as a posterior sample. Samples are rejected until the likelihood boundary is respected, though this occurs less frequently than when naively rejection sampling the prior.

The use of multiple ellipses when sampling has another distinct advantage in that it naturally enables efficient sampling of multimodal posteriors, since each mode is assigned a separate ellipse while low probability regions between these modes are avoided. Multimodality is a feature that can cause difficulties for MCMC samplers, as moving from one mode to another requires a move across the low probability region separating them. As a result, these samplers can fail to explore the full posterior distribution, instead sampling only a single mode. We expect our problem to exhibit this multimodal behaviour due to the degeneracies described in section~\ref{sec:redshift-degeneracy}, and so require a sampling method suited to this case.

The need for nested sampling methods to sample from the prior imposes some constraints on our choice of prior. MultiNest natively samples from a unit side-length hypercube and these samples are transformed into samples of the prior using a prior transform function. However, due to the discrete marginalisation over template, we cannot separate the posterior to define a prior transform function. As a result, we take the approach suggested by \cite{multinest} of defining a uniform prior to sample from, and defining the `likelihood' for MultiNest as our marginalised posterior. 

This has two main effects. Firstly, the sampling is likely to be less efficient, as the prior sampling step is not guided by the true prior, and so low-prior regions may be sampled frequently. Secondly, sampling from a uniform prior necessitates imposing a hard cut on the prior range of each parameter. Since the location of these cuts effects the value of the evidence $\mathcal{Z}$, they should not be imposed thoughtlessly. At high redshift and bright magnitudes, the priors tend to zero, meaning that the exact positions of these cuts have negligible effect on the evidence. However, this is not the case for the faint-end of the magnitude priors; setting this cut is discussed is section~\ref{sec:specify-priors}.

\subsection{\texttt{blendz} package} \label{sec:blendz}

We have written a Python package \texttt{blendz} to perform the redshift inference of blended sources described in {sections~\ref{sec:formalism} and \ref{sec:part-blend}}, and the identification of the number of components using model comparison described in section~\ref{sec:model-select}. The package supports analysis of blends with an arbitrary number of components using either the included or user-supplied template sets. The output of such an analysis is a set of samples from the joint posterior for each number of components considered, and an estimate of the Bayes factor for model comparison. The model comparison can then easily include a model prior through multiplication of the Bayes factor.

The package is also written in an object-orientated way, allowing the user to easily redefine the priors. While the supplied prior is used in this work with galaxies of either early, late or irregular types, it is written to be calibrated and used with any number of possible types. For blended sources of more then two components, the excess probability term $\xi_{\chi}^{(N)}$ is defined recursively to use the correct combination of two-point terms and assumes higher order correlations are negligible.

Documentation and instructions for installation can be found at \url{http://blendz.readthedocs.io}. The package can also be immediately installed from the official Python Package Index\footnote{\url{https://pypi.org/}} by using the \texttt{pip install blendz} command. Finally, the source is available in a \texttt{git} repository hosted at \url{https://github.com/danmichaeljones/blendz}.

\section{Results from mock observations} \label{sec:sim-results}

\subsection{Fully-blended sources} \label{sec:full-sim-results}

\begin{figure}
	\includegraphics[width=\columnwidth]{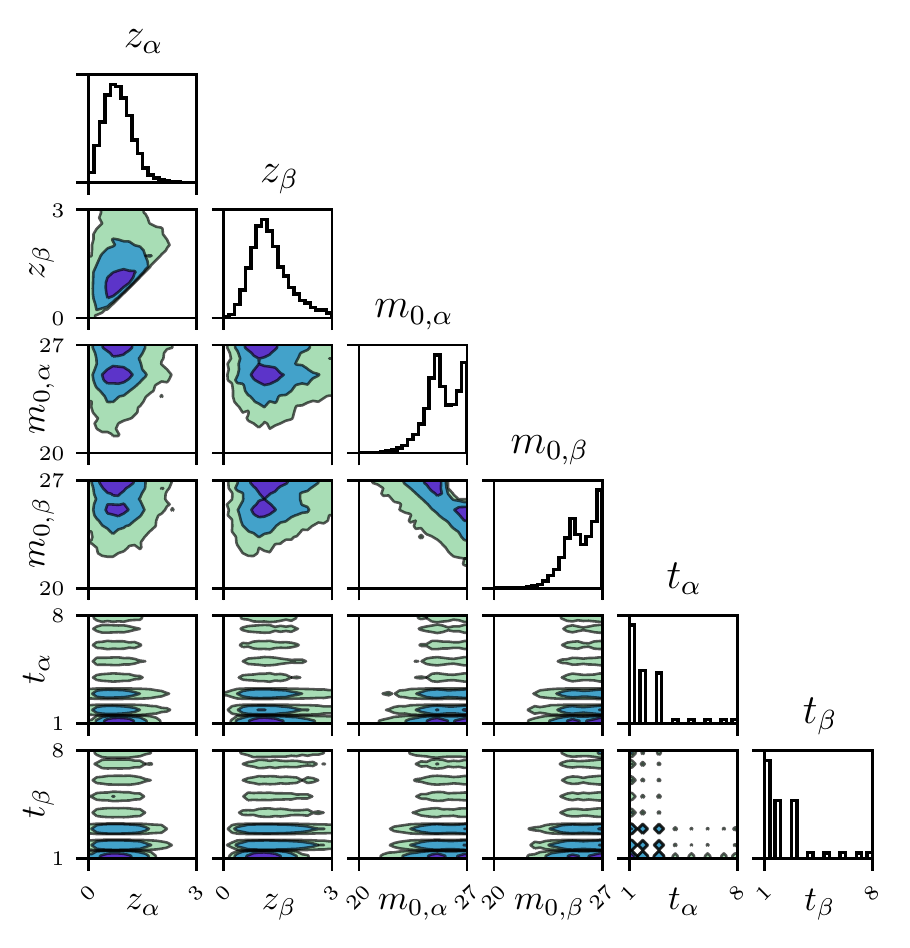}
	\caption{Corner plot of the prior sampled to create the mock catalogue. As described in the text, the bimodal shape of the marginal magnitude distributions is a result of both the selection effect and sorting components by redshift. The redshift sorting condition can be seen as a hard diagonal cut in the joint redshift distribution.}
	\label{fig:sim_distributions}
\end{figure}

As an initial test of the method, we used a Monte Carlo simulation to create a set of mock photometric observations to test our method against. These mock observations simulate an optical survey using the six LSST optical filters $u,g,r,i,z,Y$~\citep{lsstsciencebook}, with an $r$-band magnitude selection of $m_{\textrm{lim}} = 24$. We also applied hard cuts to the component magnitudes of $m_{\textrm{min}}=19$ and $m_{\textrm{max}}=26$. We then generated 1000 sources, each of which is a blend of two components in all bands. This was done by sampling a prior describing this distribution of objects using the Markov Chain Monte Carlo (MCMC) sampler \texttt{emcee}~\citep{emcee} to generate the true parameters $\{z\}, \{t\}, \{m_0\}$ for each simulated source. A plot of this prior distribution, plotted using \texttt{corner.py}~\citep{corner}, is shown in Fig.~\ref{fig:sim_distributions}. 

The effect of the selection function and the faint-end magnitude cut can be seen clearly in the two-peaked shape of the marginal distributions of $m_{0, \alpha}$ and $m_{0, \beta}$. The brighter-magnitude peak is a result of the selection function. In the single-component case, this would cause the prior to tend to zero at faint magnitudes. In the two-component case however, the magnitude priors of each component extend beyond $m_{\textrm{lim}}$ as the selection effect is applied to the combined magnitude of both components. The brighter component in a two-component blend must, by definition, contribute at least half of the total flux. As a result, the selection effect prevents the magnitude of this component from being too faint. Since we impose the sorting condition on redshifts, and the brightest component in a source is not exclusively the lower-redshift one, this action of the sorting condition causes the brighter peak in the marginal distributions of both $m_{0, \alpha}$ and $m_{0, \beta}$. If we instead impose the sorting condition on the magnitudes, these distributions become unimodal. Fig.~\ref{fig:sim_distributions} also shows the effect of the redshift sorting condition in the $(z_\alpha, z_\beta)$ marginal distribution as a hard diagonal cut.

The model fluxes for these sampled parameters were then generated using the template responses defined in section~\ref{sec:flux}. We use the template set of \cite{coeTemplates}, containing one early type, two late type and one irregular type templates from \cite{colemanTemplates}, two starburst templates from \cite{kinneyTemplates} and two starburst templates from \cite{coeTemplates}. \resp{This same template set is then used during the inference. This allows a test of the method without the effect of unrepresentative templates, a source of error that is not unique to the case of blended sources.}

Finally, we add an observational error to each observation. The flux error in band $b$ is randomly drawn from an uncorrelated, zero-centred normal distribution $\sigma_{b} \sim \mathcal{N}\left( \sigma_{b} \,|\, 0, \,\Sigma \right)$. \resp{The noise is set for all sources to be the final $1 \sigma$ depth expected from LSST~\citep{lsstSummary}.} We use these noisy observations to draw samples from the one- and two-component posteriors to test both the redshift determination and model comparison performance, setting the prior to the true distribution the photometry was sampled from.

  \begin{figure*}
  	\begin{center}$
  		\begin{array}{lll}
  		\includegraphics[width=.5\textwidth]{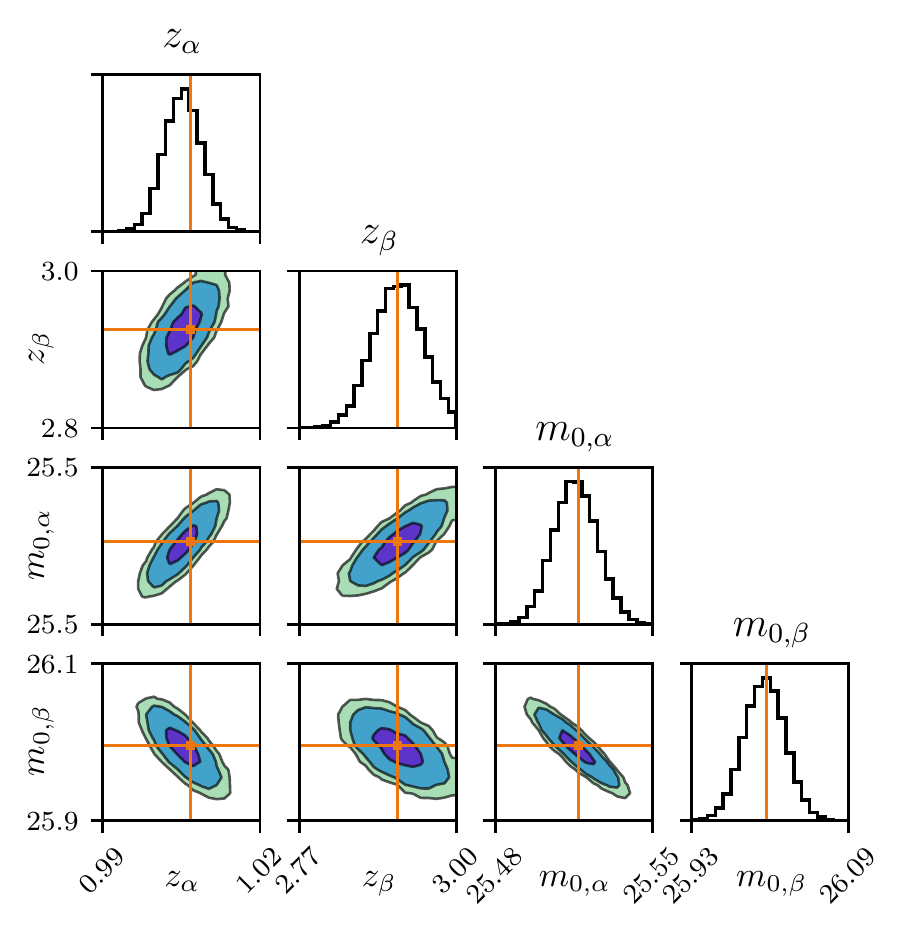} 
  		\includegraphics[width=.5\textwidth]{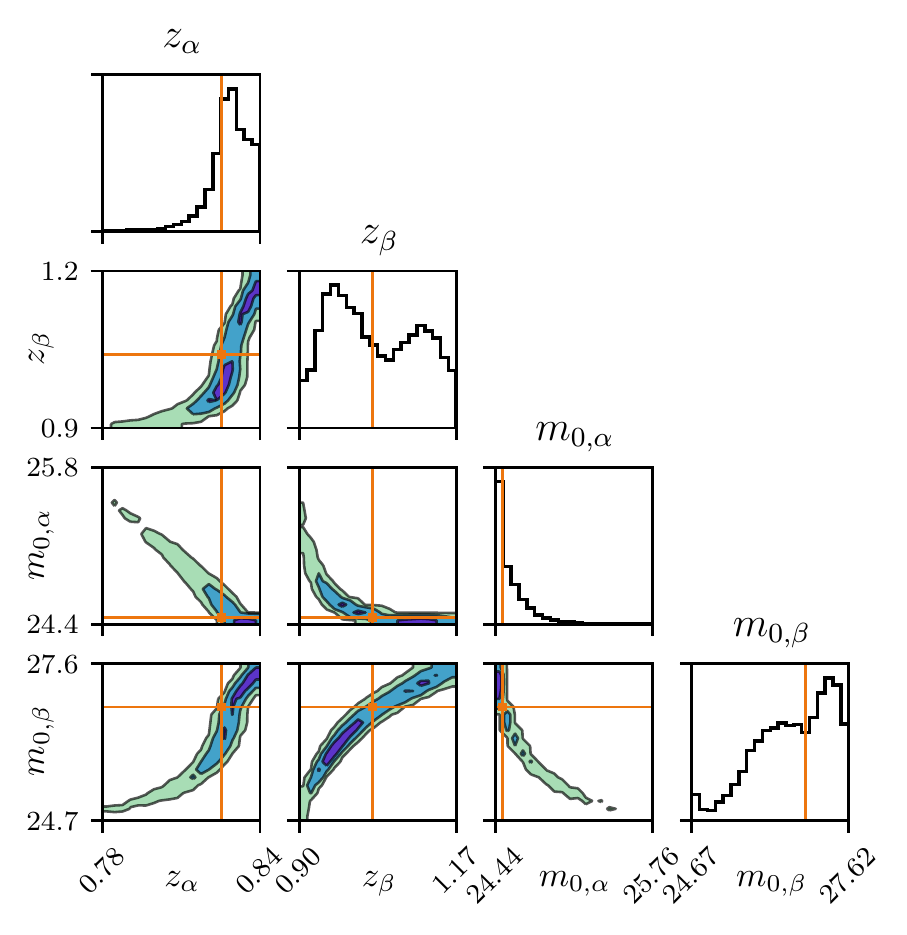} 
  		\end{array}$
  	\end{center}
  	\caption{The 4D posterior distribution output from our method for two example sources. The true parameter values are shown in orange. The left panel shows a well constrained source with some correlations between components, though the true redshift is well recovered. The right panel shows an example of a bimodal posterior that can arise in photometric redshift problems.}
  	\label{fig:many_blend_corners}
  \end{figure*}

Fig.~\ref{fig:many_blend_corners} shows two examples of the 4D posterior that is the output from our method for each sample. For plotting purposes, the number of live points used for sampling is larger than that used for the inference and model comparison results throughout this paper. However, the change in the results is negligible. The left panel shows an example of a well constrained source with a unimodal posterior. This posterior shows correlations between the component parameters; this is expected, since the total flux of each band that is well constrained by the observations is split between the components. Reducing the model flux in a band of one component will result in a compensation in the other component, correlating their parameters.

The right panel of Fig.~\ref{fig:many_blend_corners} shows a particularly prominent example of the curved degeneracies that can arise in the blended posteriors. This is due to the total magnitude of the source being well constrained by noisy observations, while component magnitudes are not themselves observable. This leads to a degeneracy that is curved due to the non-linearity of adding magnitudes. A result of this curved degeneracy is bimodality in the marginalised posterior of $z_\beta$. However, there is still significant probability density around the true redshift, highlighting the importance of not compressing the information content of a full posterior distribution into only a small set of numbers.

   \begin{figure*}
   	\includegraphics[width=\textwidth]{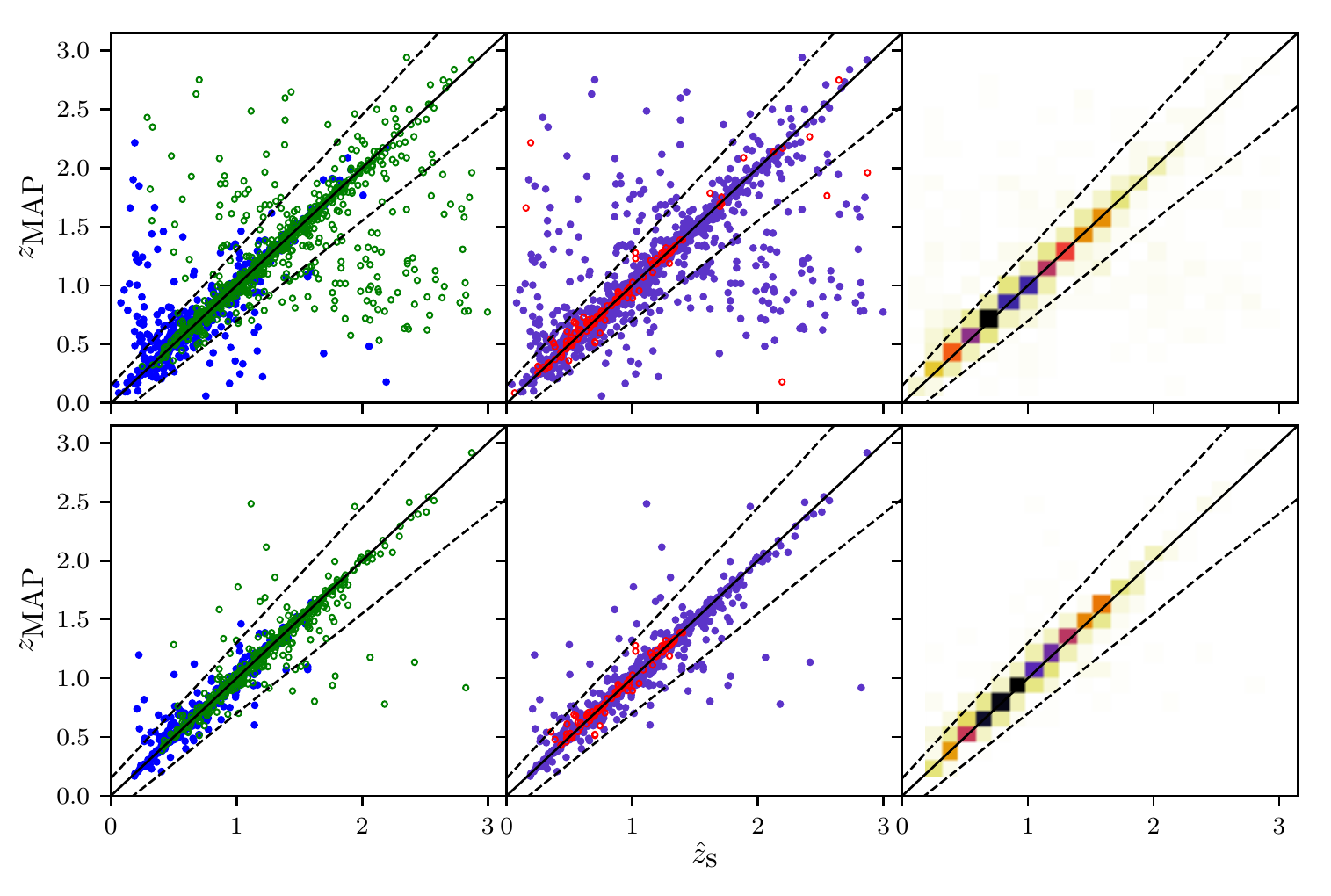}
   	\caption{Scatter plot comparing the maximum \textit{a posteriori} point estimates from the photometric redshift estimation with the true redshifts for the mock observations. The left panels distinguish the components, with $z_\alpha$ plotted with closed blue markers, and $z_\beta$ plotted with open green markers. The centre panels show the blend identification, with sources identified as blends plotted with closed purple markers, and those misidentified as single sources plotted with open red markers. The right panels show a 2D histogram of the combined sample. Panels in the top row show the results for the full mock catalogue, while the bottom row only includes sources where the standard deviation of samples from each redshift marginal-posterior are sufficiently small, $\sigma_\alpha \leq 0.2 \,\forall\, \alpha$. The dashed lines in each panel show an error of $0.15(1 + z)$. }
   	\label{fig:sim-results}
   \end{figure*}

Fig.~\ref{fig:sim-results} shows a comparison of the photometric estimation of the component redshifts against their true simulated values. Point estimates of the redshift $z_\textrm{MAP}$ are obtained by taking the maximum \textit{a posteriori} (MAP) value of each component redshift posterior, marginalising over the other three parameters. The method recovers the true redshift of each component from simulated photometry well. The performance of photometric redshift methods is often summarised by the RMS scatter $\sigma_{\textrm{RMS}}$. We first define the normalised error for galaxy $g$ as
\begin{equation}
\tilde \delta z_g = \frac{\hat z_{s, g} - z_{\textrm{MAP}, g}}{1+ \hat z_{s, g}} \,,
\end{equation}
where each galaxy $g$ is a single component of a blended source. Writing the total number of galaxies in our test catalogue as $N_{g}$, we then define the RMS scatter as
\begin{equation}
\label{eqn:rms_scatter}
\sigma_{\textrm{RMS}} = \sqrt{\frac{1}{N_{g}} \sum_{g} \left(\tilde \delta z_g \right)^2 } \,.
\end{equation}

Computing this quantity for our mock blended observations, we find an RMS scatter of $\sigma_{\textrm{RMS}} = 0.163$. \resp{This compares to a scatter of $\sigma_{\textrm{RMS}} = 0.0267$ for photometric redshifts of mock observations of single sources, demonstrating the added difficulty of blending. }

This scatter can be improved by excluding sources with photometric redshifts that, using the uncertainty information of the posterior distribution, are identified as untrustworthy. This is done by comparing a summary statistic against a threshold that controls the stringency of the test; we use the standard deviation of redshift marginal-posterior samples $\sigma_\alpha$ separately for each component, though a variety of summary statistics are available. Keeping only sources with $\sigma_\alpha \leq 0.2 \,\forall\, \alpha$, the RMS scatter is reduced to $\sigma_{\textrm{RMS}} = 0.064$, with $37\%$ of sources removed. The effect of this is shown in the bottom row of Fig.~\ref{fig:sim-results}.

\resp{The percentage of outliers can also be quantified. Outliers are defined as sources where either component has an error $|z_{\textrm{MAP}} - \hat z_s| \geq 0.15 (1 + \hat z_s)$. For the full set of mock observations, this percentage was found to be $18.6\%$. By keeping only sources with $\sigma_\alpha \leq 0.2 \,\forall\, \alpha$ as described above, the percentage of outliers falls to only $6.0\%$.}

The results of the detection of blends are also shown in the centre panels of Fig.~\ref{fig:sim-results}. By using equation~\ref{eqn:model-select}, we calculate $\mathcal{P}_{2, 1}$, the relative probability that a source is a two-component blend compared to a single source. The interpretation of this probability is problem-dependent; a probability of $\ln \left( \mathcal{P}_{2, 1} \right) > 0$ indicates a preference towards the source being blended, while a threshold to $\ln \left( \mathcal{P}_{2, 1} \right) > 5$ indicates strong evidence~\citep{bayesfact}. Likewise, probabilities of $\ln \left( \mathcal{P}_{2, 1} \right) < 0$ and $\ln \left( \mathcal{P}_{2, 1} \right) < -5$ indicate a preference and strong evidence for the single source case respectively. As the blended and single thresholds are pushed more positive and negative respectively, there are sources with values of $\ln \left( \mathcal{P}_{2, 1} \right)$ that fall between these thresholds. In these cases, the source is assigned neither label.

\begin{figure*}
	\includegraphics[width=\textwidth]{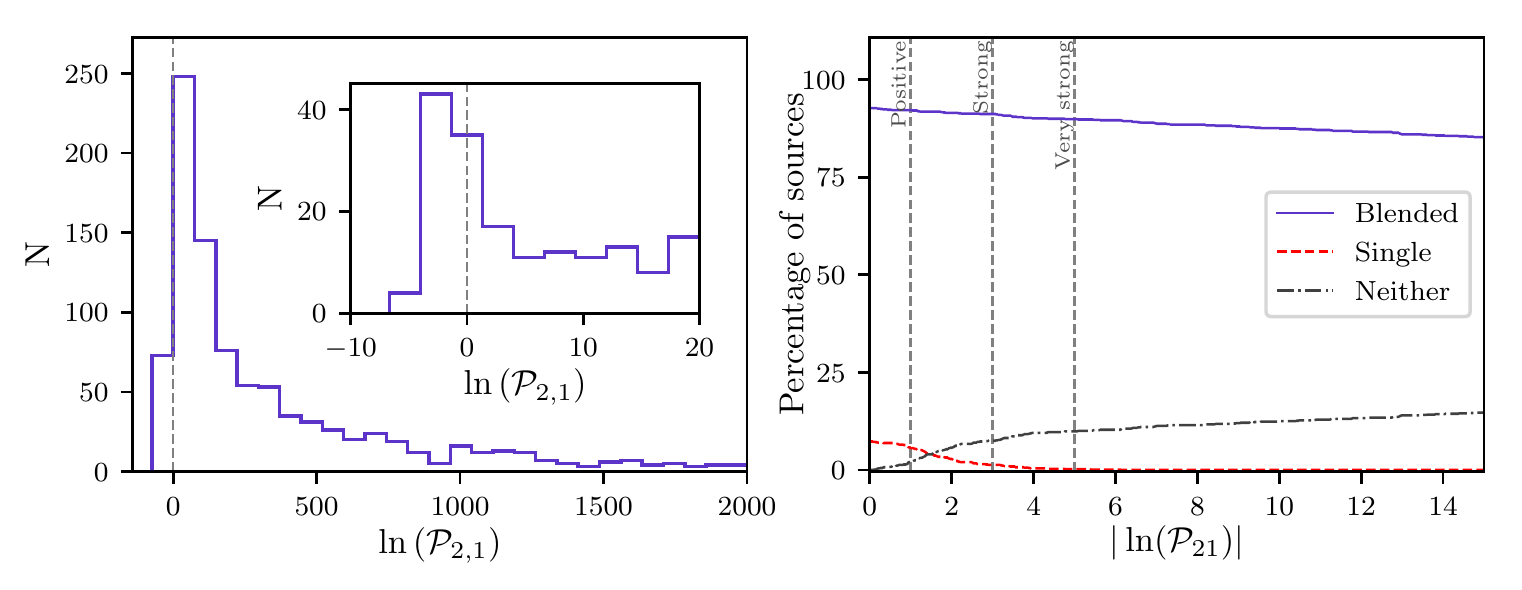}
	\caption{The left panel shows the distribution of the relative blend-to-single probability for the mock catalogue, with the inset showing the same distribution, zoomed around lower relative probabilities {and binned more finely}. The right panel shows the percentage of sources assigned as either blended, single sources or not assigned to either as the threshold for deciding between each label is changed.}
	\label{fig:sim_bayes_hist}
\end{figure*}

As described in section~\ref{sec:model-select}, we assume the relative prior probability of a blend to be $P(N = 2) / P(N = 1) = 1$, i.e., we give no preference to either model. Under this assumption, the method identifies $92.7\%$ of sources as blends and $7.3\%$ as single sources. Increasing the threshold to strong evidence, we find that $89.9\%$ of sources are identified as blends and $0.2\%$ as single sources; the remaining $9.9\%$ fall between these thresholds. The distribution of the relative probability of blending and the effect on blend identification of changing the threshold are shown in Fig.~\ref{fig:sim_bayes_hist}.

These results show that the method can both recover the redshifts from broadband observations of blended objects, and detect the blending of a large fraction of these objects from their photometry alone. In addition, the output from these tests are not just point estimates of redshifts, but the full four-dimensional posterior distributions that capture the correlations between components that can be lost by working with component separated maps. These are the results of simulated observations, however; real data has the complication that the flux model is no longer exact, i.e., the templates are not perfectly representative of all galaxies observed. As such, we test the method on real data in section~\ref{sec:gama-data}.

\subsection{\resp{Partially-blended sources}} \label{sec:part-sim-results}

\resp{
To test the effect of adding resolved data, we created a set of mock photometric observations of two-component partially blended systems. These observations simulate the same six-band optical survey as described in section~\ref{sec:full-sim-results}, combined with a four-band optical and infrared space-based survey using the Euclid filters $vis,\, Y, J, H$~\citep{euclidDesign}. This latter survey is assumed to have made resolved measurements of each component, while the former is fully-blended as before. Thus, our partially-blended data vector contains $14$ fluxes for each source.}

\resp{
For comparison with the partially-blended results, we repeat the inference several times. Firstly, we compare against the fully-blended LSST-like case described above. Next, we compare to an inference using the resolved Euclid bands only, testing the effect of removing the difficulty of blending but using lower signal-to-noise data. Finally, we test against the case of using both the LSST- and Euclid-like data, but assuming that sources are blended in all bands. This allows us to separate the improvement as a result of adding resolved data from that of simply having more bands available.}

\resp{
For the fully-blended bands, we reuse the simulated fluxes described in section~\ref{sec:full-sim-results}. For the resolved bands, we generate observed fluxes using the same randomly sampled source parameters. The observed fluxes are then generated using the flux model described in section~\ref{sec:flux} with added observational errors drawn randomly from an uncorrelated, zero centred normal distribution. The noise in these resolved bands is set to the final $1 \sigma$ depths expected from Euclid observations~\citep{euclidSummary}.}

\resp{
We use the same prior as described in section~\ref{sec:full-sim-results} for both the simulation and inference steps. The reference band over which the prior is defined is set to be the $r$-band of the blended observations. This band is not present during the inference step using only the resolved data. As a result, we use the flux model from section~\ref{sec:flux} to convert between $r$- and $Y$-band magnitudes before evaluating the prior.}

\begin{figure*}
	\includegraphics[width=\textwidth]{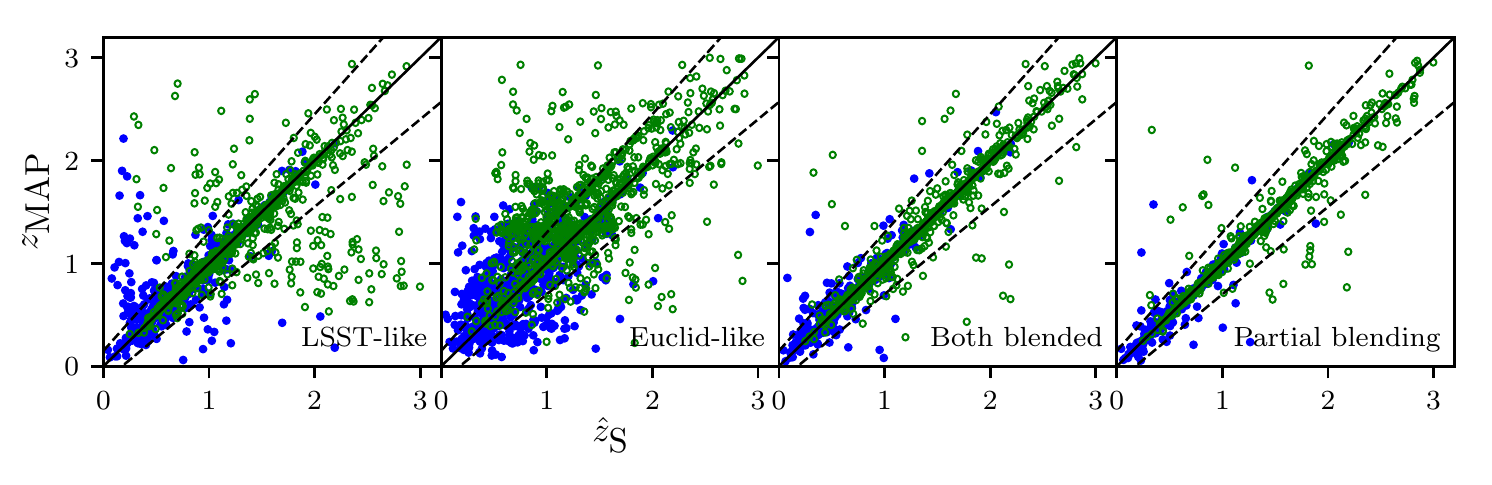}
	\caption{\resp{Scatter plot comparing the maximum \textit{a posteriori} point estimates for the fully-blended, resolved and partially-blended cases. The closed blue markers represent the redshift of the closer component, $z_\alpha$, while the open green markers represent the redshift of the more distant component, $z_\beta$.}}
	\label{fig:part-sim-scatter}
\end{figure*}

\resp{
Fig.~\ref{fig:part-sim-scatter} shows a comparison of the photometric redshift point estimates with the true simulated values for the four sets of inferences. As before, these point estimates are the MAP values of the marginal redshift posteriors.}

\resp{
The left panel shows the fully blended case, the same results as section~\ref{sec:full-sim-results}. Again, the redshifts of many components are well recovered, though there is a significant fraction of outliers. The RMS scatter in this case was found to be $\sigma_{\textrm{RMS}} = 0.163$. The percentage of outliers, sources where either component has an error $|z_{\textrm{MAP}} - \hat z_s| \geq 0.15 (1 + \hat z_s)$, was found to be $18.6\%$.
}

\resp{
The centre-left panel of Fig.~\ref{fig:part-sim-scatter} shows the results for the resolved observations. Though finding the photometric redshift of resolved sources is an easier inference problem, this is counteracted by the significant reduction in the signal-to-noise of this data. As a result, we find an RMS scatter of $\sigma_{\textrm{RMS}} = 0.212$ with $55.0\%$ of sources marked as outliers.}

\resp{
The centre-right panel of  Fig.~\ref{fig:part-sim-scatter} shows the results for combination of LSST- and Euclid-like data in the fully blended case. We find that the addition of the four Euclid bands significantly improves the precision of the redshift inference, which has an RMS scatter of $\sigma_{\textrm{RMS}} = 0.073$. The fraction of outliers has also improved significantly, with only $6.6\%$ of sources marked as outliers.}

\resp{
Finally, the right panel of Fig.~\ref{fig:part-sim-scatter} shows the results for the partially blended case, combining the high-precision blended observations with the resolved data. Here, we find that the RMS scatter has reduced to $\sigma_{\textrm{RMS}} = 0.065$, a factor of $2.5$ improvement over the blended LSST-like data alone, and a factor of $1.12$ over the combined LSST- and Euclid-like blended data. The percentage of outliers has also been reduced. Here, only $3.4\%$ of sources are found to be outliers, a factor of $5$ improvement over the fully-blended case, and a factor of $1.9$ improvement over the combined LSST- and Euclid-like blended data.}

\resp{
While the most significant improvement was obtained through the increase in the number of bands, these results show that the quality of photometric redshifts of blended sources can be improved through the inclusion of resolved data. This is particularly apparent in the reduction of outliers. One advantage conferred by the addition of resolved data is a constraint on the relative magnitudes of blended components. In the fully-blended case, the reference-band magnitude of each of the components must be inferred from the combined magnitude of the blended source only. This can lead to the degenerate distributions shown in Fig.~\ref{fig:many_blend_corners}. Adding resolved photometry can help to break this degeneracy by providing information about each component individually. The precision of the photometric redshift inferences is therefore improved.}

\begin{figure*}
	\begin{center}$
		\begin{array}{lll}
		\includegraphics[width=.5\textwidth]{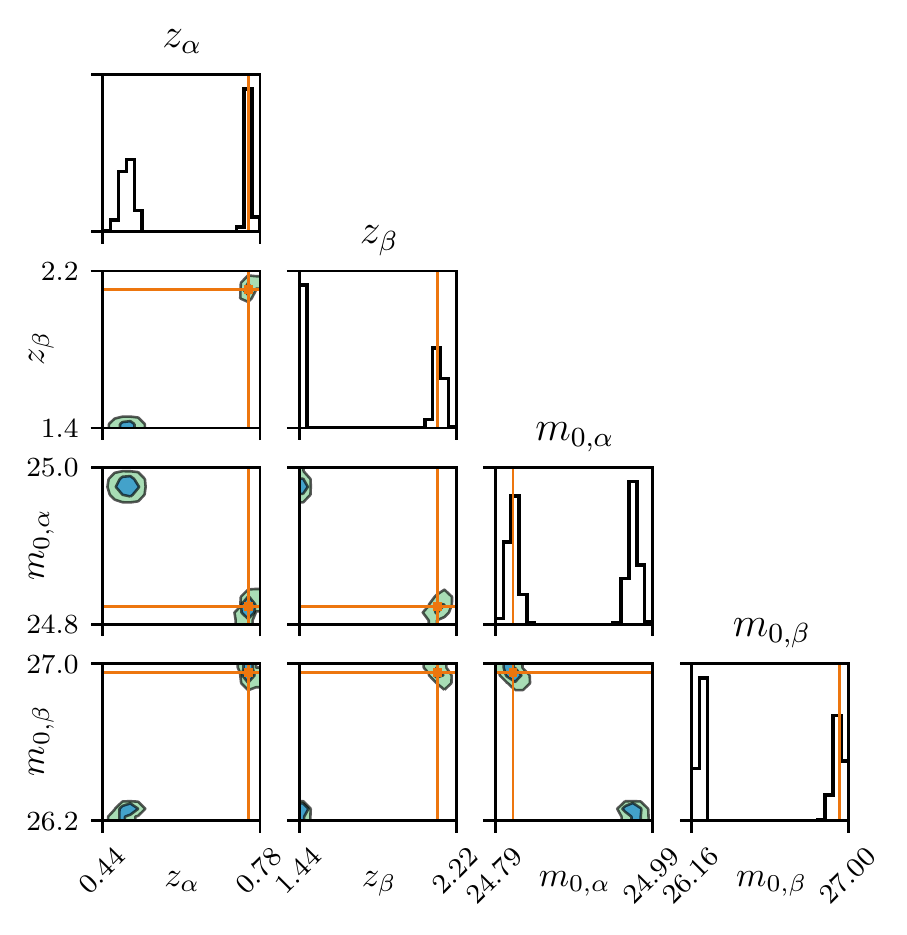} 
		\includegraphics[width=.5\textwidth]{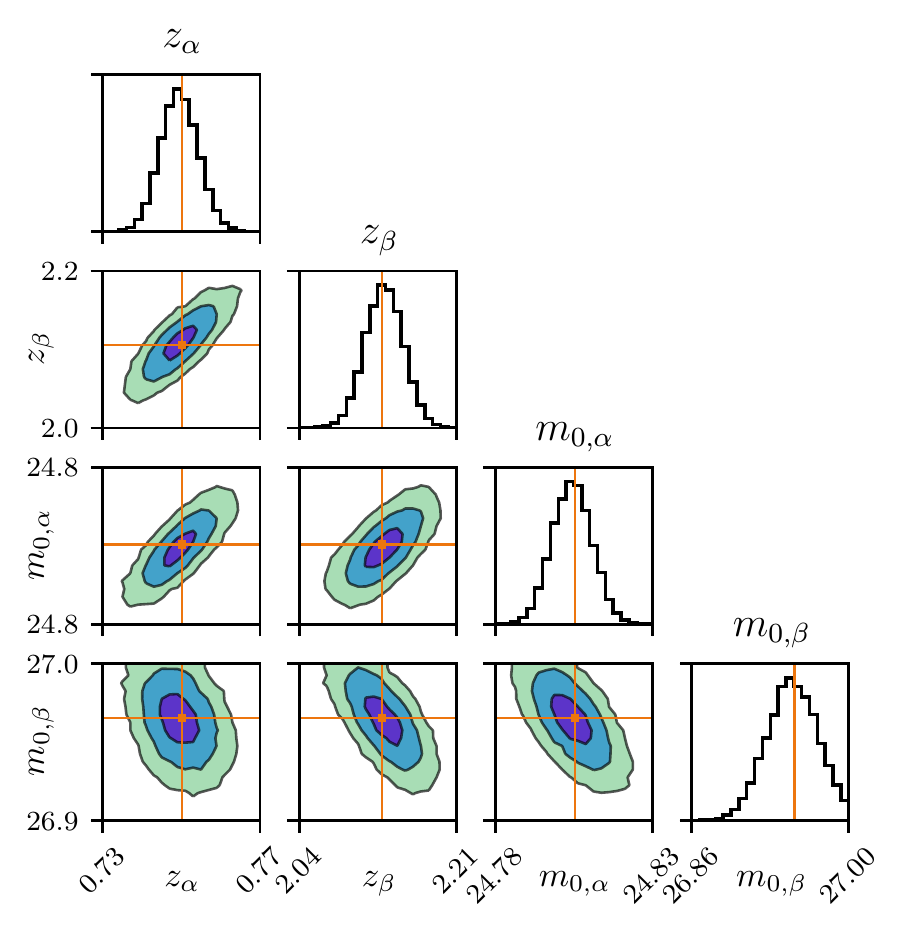} 
		\end{array}$
	\end{center}
	\caption{\resp{The 4D posterior distributions for a two-component blended source in the fully-blended and partially-blended cases. The left plot shows the result of inference using blended data only. While their is significant posterior density around the true parameter values shown by the orange line, this posterior is highly bimodal, with two distinct solutions that cannot be distinguished. The plot on the right shows the result of the partially-blended case that includes both blended and resolved observations. The addition of information about the magnitude of each component separately has removed the incorrect mode, resulting in a posterior that recovers the true solution well.}}
	\label{fig:part-blend-corner}
\end{figure*}

\resp{
An example of this phenomenon is shown in Fig.~\ref{fig:part-blend-corner}. The left panel shows a corner plot of the posterior distribution for a fully-blended source. The marginal distributions for each component redshift are highly multimodal, with well separated redshifts occurring at distinct magnitudes. Though there is significant posterior density around the true redshifts, the MAP point estimate of $z_\beta$ would show a significant error, as the incorrect mode has a higher posterior. The right panel of Fig.~\ref{fig:part-blend-corner} shows the same source analysed in the partially-blended case after the addition of the resolved photometry. Here, the width of the posterior has been significantly reduced by the removal of the incorrect mode. The posterior now shows that the redshift of the source has been well constrained, and the redshift point estimates would no longer have a large error.}

\section{GAMA blended sources catalogue}
\label{sec:gama-data}

The Galaxy And Mass Assembly (GAMA) survey~\citep{gamaData} is a spectroscopic galaxy survey that observed $286 \textrm{ deg}^2$ of sky over several regions to a magnitude limit of between $r<19$ and $r<19.8$. In doing so, it obtained precise redshifts of $>150 \, 000$ sources. The observed regions were chosen to overlap with existing imaging surveys such as Sloan Digital Sky Survey (SDSS)~\citep{sdssData} and VISTA Kilo-degree Infrared Galaxy (VIKING) Survey~\citep{vikingData}. As a result, the spectroscopic data is accompanied by a set of aperture-matched photometry covering nine filter bands $u,g,r,i,z,Y,J,H,K$ from optical to infrared wavelengths~\citep{gamaPhotometry}.

The GAMA blended sources catalogue~\citep{gamaBlends} contains 280 sources from the GAMA survey that have been spectroscopically identified as blended objects. These were selected using an automated template-based spectrum fitting method~\citep{gamaAutoz} that cross correlates galaxy templates with the observed spectra to determine the galaxy redshift. Sources where two different redshifts showed strong cross-correlations were visually inspected, resulting in a selection of blended galaxies. The motivation of \cite{gamaBlends} was the identification of strong lens candidates. However, a catalogue of spectroscopically identified blended galaxies with accompanying nine-band photometry gives us an useful test case for the blended photometric redshift estimation method on non-simulated photometry with secure redshifts available for both components.

We first calibrate the prior using the procedure described in section~\ref{sec:calibrate-priors}. To do this, we used $26782$ unblended, well-observed galaxies. These were selected by enforcing every band to be free from SExtractor~\citep{sextractor} error flags and excluding all galaxies in the blended source catalogue. The resulting prior from the calibration procedure is shown in Fig.~\ref{fig:gama-prior-calibration}. As discussed in section~\ref{sec:specify-priors}, we test two methods of setting the faint-end magnitude cut $m_{\textrm{max}}$, firstly as a $5\sigma_0$ flux deviation using equation~\ref{eqn:mag-max-sigma}, and secondly, fixing $m_{\textrm{max}}=20.8$. Throughout, we refer to these as the sigma-$m_{\textrm{max}}$ case and fixed-$m_{\textrm{max}}$ case respectively. \resp{We then proceed with the inference using the same template set\footnote{\resp{The templates used to fit the spectroscopic redshifts as described in~\cite{gamaBlends} do not cover the full wavelength range of the photometry. As a result, we do not use them for the photometric redshift inference.}}  described in section~\ref{sec:sim-results}.}

\begin{figure*}
	\includegraphics[width=\textwidth]{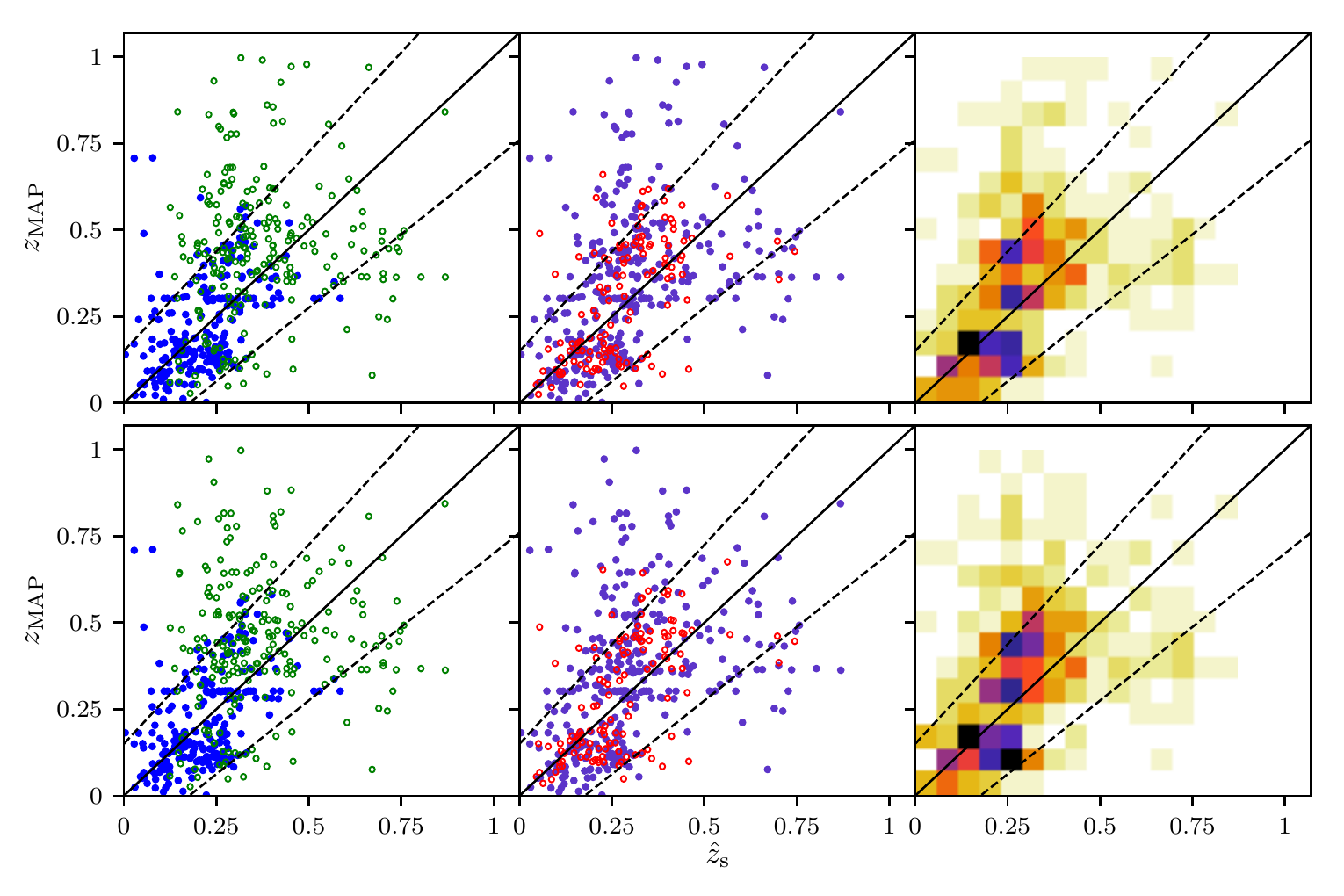}
	\caption{
		Scatter plot comparing the maximum \textit{a posteriori} point estimates from the photometric redshift estimation with the spectroscopic redshifts for sources from the GAMA blended sources catalogue. The left panels distinguish the components, with $z_\alpha$ plotted with closed blue markers, and $z_\beta$ plotted with open green markers. The centre panels show the blend identification, with sources identified as blends plotted with closed purple markers, and those misidentified as single sources plotted with open red markers. The right panels show a 2D histogram of the combined sample. Panels in the top row show the results for the 
		sigma-$m_{\textrm{max}}$ case, while those in the bottom row show the fixed-$m_{\textrm{max}}$ case. The dashed lines in each panel show an error of $0.15(1 + z)$.}
	\label{fig:gama-results}
\end{figure*}

The resulting redshift point estimates are shown in Fig.~\ref{fig:gama-results}. While noisier than the simulated case, the method still recovers reasonable estimates; using equation~\ref{eqn:rms_scatter}, we find an RMS scatter of $\sigma_{\textrm{RMS}} = 0.156$ in the both the sigma-$m_{\textrm{max}}$ and fixed-$m_{\textrm{max}}$ cases. \resp{This compares to the scatter for a set of unblended GAMA sources of $\sigma_{\textrm{RMS}} = 0.116$. Obtaining this value even without the added complication of blending suggests a mismatch between the sources and the template set.}

\begin{figure*}
	\includegraphics[width=\textwidth]{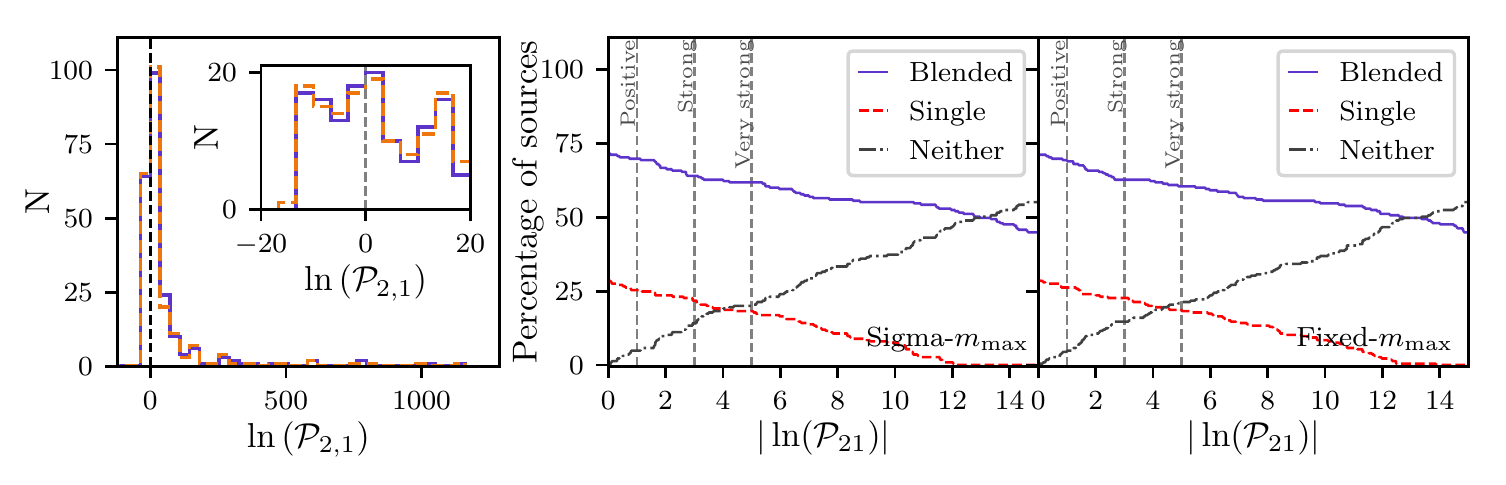}
	\caption{Plots showing the differences in the model comparison results between the two methods tested of setting the faint-end magnitude cut $m_{\textrm{max}}$, labelled the sigma-$m_{\textrm{max}}$ and fixed-$m_{\textrm{max}}$ cases. The left panel shows the distribution of the relative blend-to-single probability for the mock catalogue, with the inset showing the same distribution, zoomed around lower relative probabilities {and binned more finely}. The solid line shows the sigma-$m_{\textrm{max}}$ case, and the dashed line shows the fixed-$m_{\textrm{max}}$ case. The two right panels shows the percentage of sources assigned as either blended, single sources or not assigned to either as the threshold for deciding between each label is changed. }
	\label{fig:gama_bayes_hist}
\end{figure*}

We also compute the inferred blend probability $\mathcal{P}_{2, 1}$ for these galaxies. The distribution of these probabilities is shown in Fig.~\ref{fig:gama_bayes_hist}. As described in section~\ref{sec:sim-results}, $\ln\left(\mathcal{P}_{2, 1}\right) > 0$ and $\ln\left(\mathcal{P}_{2, 1}\right) > 5$ show a preference and strong evidence for a blended source respectively, while $\ln\left(\mathcal{P}_{2, 1}\right) < 0$ and $\ln\left(\mathcal{P}_{2, 1}\right) < -5$ show the same for the single source case. The distribution of the blend probability and the effect of the evidence threshold on blend identification is shown in  Fig.~\ref{fig:gama_bayes_hist}.

In our tests of the sigma-$m_{\textrm{max}}$ case, $71.6\%$ of sources showed a preference for being blended, with $28.4\%$ preferring a single source. Increasing the threshold to strong evidence, these percentages fall to $61.8\%$ and  $18.2\%$ respectively. Finally, the incorrectly identified single sources can be excluded entirely by increasing the threshold to $\left|\ln\left(\mathcal{P}_{2, 1}\right)\right| < 12.5$, with  $50.7\%$ of sources identified as blends at this level. 

The identification of blends was very similar in the fixed-$m_{\textrm{max}}$ case. We found that $71.1\%$ of sources showed a preference for being blended, and $28.9\%$ preferred a single source. At the strong evidence threshold, $60.4\%$ of sources are correctly identified as blends, with $18.2\%$ misidentified as single sources. The threshold to exclude misidentified sources completely in the fixed-$m_{\textrm{max}}$ case is $\left|\ln\left(\mathcal{P}_{2, 1}\right)\right| < 13.9$, slightly higher than the sigma-$m_{\textrm{max}}$ case. At this level, $48.0\%$ of sources are still correctly identified as blends.

These results show that photometric redshift estimates can be obtained for blended sources, and that the method can identify many blended sources from just their broadband photometry. By adjusting the threshold of the probability $\mathcal{P}_{2, 1}$, blended sources can be selected in a way that trades off completeness and purity.

Several techniques for improving the scatter of photometric redshifts have been proposed, such as rest-frame template error functions~\citep{eazy}, iterative methods to modify templates to be more representative~\citep{zebra}, using clustering-based redshift estimation to calibrate systematic biases using galaxies~\citep{desBias} and intensity mapping observations~\citep{intensitymap}, and constructing priors in terms of physical galaxy properties~\citep{physicalpriors}. While an investigation of these methods is beyond the scope of this paper, they could also be applied while using this method. This could help to reduce the scatter of the blended photometric redshift estimates to a level necessary for future surveys, while retaining the full information of the posterior for accurate error propagation.


\section{Conclusions}
\label{sec:conclusions}

Blended sources will become far more common in future galaxy surveys than are found currently due to increases in the depth of photometry and as a result, the number density of galaxies. We present a Bayesian photometric redshift method that generalises the existing BPZ~\citep{bpz} method to the case of blended observations. We derive a posterior for the redshift and magnitude of each component which we sample to obtain estimates of the redshift. We also use this posterior in a model comparison procedure to infer the number of components in a source.

By doing this, the method is able to infer both the redshift of each component within a blended source, and identify that a source is blended from its broadband photometry alone. The joint posterior distribution of the redshifts of all components in a blend provides a complete accounting of the correlations in the final result, information that can be lost when separating components and estimating redshifts for each component separately. This uncertainty information is essential for obtaining accurate uncertainties on cosmological parameters that rely on the photometric redshift estimates. A Python implementation of the method, \texttt{blendz}, is available to download.

\resp{
By inferring the redshifts of components directly from their blended photometry, the method presented here is directly applicable to ambiguously blended objects that cannot otherwise be deblended. The partial-blending formalism described in section~\ref{sec:part-blend} also enables the catalogue-level joint analysis of sources in space- and ground-based surveys such as Euclid and LSST. The complementarity of these surveys will allow cosmological parameters to be constrained more precisely than either survey could individually, and analysis of blended sources from their aperture photometry will be simpler than a joint pixel-level analysis~\citep{synergy}. 
}

\resp{
The method presented here could also be combined with existing deblending methods that utilise the spatial information of images directly. These methods are complementary; image-based deblending methods are effective provided that components are sufficiently well separated. If this is not the case, there is too little spatial information to be able to separate components, and colour information is necessary. Combining these methods could allow future surveys to identify a greater proportion of blended sources, reducing their effects on cosmological constraints. Deblending methods that also incorporate colour information would need to be combined with this method more carefully however, as the colour information would be used twice and thus the blending probabilities would not be independent. This method could instead be extended to incorporate imaging data by constructing a forward model of the galaxy in each band and constraining both morphology and redshift simultaneously.
}

\section*{Acknowledgements}
We thank Andrew Jaffe, Daniel Mortlock and Josh Greenslade for helpful discussions, \resp{and the referee Michael Troxel for helpful comments.}
%
DMJ acknowledges funding from STFC through training grant ST/N504336/1.
GAMA is a joint European-Australasian project based around a spectroscopic campaign using the Anglo-Australian Telescope. The GAMA input catalogue is based on data taken from the Sloan Digital Sky Survey and the UKIRT Infrared Deep Sky Survey. Complementary imaging of the GAMA regions is being obtained by a number of independent survey programmes including GALEX MIS, VST KiDS, VISTA VIKING, WISE, Herschel-ATLAS, GMRT and ASKAP providing UV to radio coverage. GAMA is funded by the STFC (UK), the ARC (Australia), the AAO, and the participating institutions. The GAMA website is http://www.gama-survey.org/.
Based on observations made with ESO Telescopes at the La Silla Paranal Observatory under programme ID 179.A-2004. 






\bsp	
\label{lastpage}
\end{document}